\documentclass[a4paper,english,useAMS,usenatbib]{mn2e}
\usepackage[totalwidth=480pt,totalheight=680pt,layoutvoffset=0.5cm]{geometry}
\usepackage[latin9]{inputenc}
\usepackage{color}
\usepackage{float}
\usepackage{url}%
\usepackage{amsmath}
\usepackage{amssymb}
\usepackage{graphicx}
\usepackage{times} 
\usepackage{fixltx2e}

\usepackage[LGR,T1]{fontenc}
\usepackage{esint}
\usepackage[authoryear]{natbib}

\makeatletter
 
\pdfpageheight\paperheight
\pdfpagewidth\paperwidth

\DeclareRobustCommand{\greektext}{%
  \fontencoding{LGR}\selectfont\def\encodingdefault{LGR}}
\DeclareRobustCommand{\textgreek}[1]{\leavevmode{\greektext #1}}
\ProvideTextCommand{\~}{LGR}[1]{\char126#1}

\usepackage{astrobib_mnras2e}

\usepackage{babel}

\newcommand{\lcdm}{$\Lambda$CDM }
\newcommand{\mpch}{h^{-1}{\rm Mpc}}
\newcommand{\msunh}{h^{-1}M_\odot}

\newcommand{\pmem}{p_{\rm mem}}

\newcommand{\avg}[1]{\langle #1 \rangle}
\newcommand{\avrmem}{\avg{R_{\rm mem}}}

\makeatother

\begin{document}

\title{On the measurements of assembly bias and splashback radius using optically selected galaxy clusters}

\author[Sunayama \& More]{Tomomi Sunayama$^{1}$\thanks{tomomi.sunayama@ipmu.jp}, Surhud More$^{1,2}$\\
$^1$Kavli IPMU (WPI), UTIAS, The University of Tokyo, Kashiwa, Chiba 277-8583, Japan\\
$^2$Inter-University Centre for Astronomy \& Astrophysics, Ganeshkhind, Post Bag 4, Pune 411007, India
}

\maketitle
\begin{abstract}
We critically examine the methodology behind the claimed observational detection
of halo assembly bias using optically selected galaxy clusters by
\citet{Miyatake2016} and \citet{More2016}. We mimic the optical
cluster detection algorithm and apply it to two different mock catalogs
generated from the Millennium simulation galaxy catalog, one in which
halo assembly bias signal is present, while the other in which the
assembly bias signal has been expressly erased. We split each of these
cluster samples into two using the average cluster-centric distance of
the member galaxies to measure the difference in the clustering
strength of the subsamples with respect to each other. We observe that
the subsamples split by cluster-centric radii show differences in
clustering strength, even in the catalog where the true assembly bias
signal was erased. We show that this is a result of contamination of
the member galaxy sample from interlopers along the line-of-sight.
This undoubtedly shows that the particular methodology adopted in the
previous studies cannot be used to claim a detection of the assembly
bias signal. We figure out the tell-tale signatures of such
contamination, and show that the observational data also shows similar
signatures. Furthermore, we also show that projection effects in
optical galaxy clusters can bias the inference of the 3-dimensional edges of
galaxy clusters (splashback radius), so appropriate care should be taken while
interpreting the splashback radius of optical clusters.
\end{abstract}
\begin{keywords} cosmology: theory \textendash{} large-scale structure of Universe \textendash{} galaxies: clusters: general. \end{keywords}

\section{Introduction}

In the \lcdm paradigm, massive dark matter halos that host clusters of
galaxies form at rare peaks of the initial density field. Such dark
matter halos are biased tracers of the matter density and are more
clustered than the underlying dark matter distribution itself. The
clustering of dark matter halos primarily depends on halo mass (e.g
\citet{Kaiser1984,Efstathiou1988,MoWhite1996}). However, there have
been several studies which show that the clustering of halos at fixed
mass can depend upon secondary properties such as the formation time
and the concentration of halos (e.g \citet{gao_etal05, wechsler06,
gao_white07, faltenbacher_white10, dalal_etal08}). This dependence of
the halo clustering amplitude on secondary properties related to their
assembly history other than the mass is commonly referred to as
\textit{halo assembly bias.} 

Even though halo assembly bias is a well-known phenomenon in
simulations and has been studied extensively, observational detection
of halo assembly bias has been difficult and controversial (e.g
\citet{Yang2006,wang_etal13,Lin2016}).  Recently, \citet{Miyatake2016}
claimed evidence of halo assembly bias on galaxy cluster scales. They
split the optically selected redMaPPer galaxy cluster catalog, using
the compactness of member galaxies in the clusters as a proxy of halo
concentration. Using weak gravitational lensing, they confirmed that
the halo masses of these cluster subsamples were consistent with each
other and yet the clustering amplitudes were significantly different.
The difference in the clustering amplitude of the two subsamples was
almost $60\%$, much larger than the difference expected from N-body
simulations.

In a follow-up study, \citet{More2016} used the cross-correlation
function of the same cluster subsamples with photometric SDSS galaxies
to confirm the difference in the clustering amplitude of these
subsamples with an enhanced signal-to-noise ratio. Additionally,
\citet{More2016} measured the location of the splashback radius for
these cluster subsamples. The splashback radius is the location of the
first turn-around of the particles and is considered as a physical
boundary of halos. The splashback radius depends on the accretion
rates of halos (e.g., \citet{diemer14,more15}). \citet{More2016} found that
the low-concentration cluster subsample had a larger splashback radius
than the high-concentration cluster subsample. Curiously, the inferred
location of the splashback radius was smaller than that expected from
numerical simulations.


There were two studies which critically examined the works by
\citet{Miyatake2016} and \citet{More2016}. \citet{Zu2017} examined
whether the difference in the clustering amplitudes of the two cluster
subsamples arose from biases in the cluster-centric radius due to
projection effects. They proposed to use only those member galaxies
which have high membership probabilities in order to define the
cluster-centric average radius of member galaxies in order to reduce
the contamination from projection effects. They showed that when such
a radius is used to define the cluster subsamples, the clustering
difference is reduced to levels which make it consistent or smaller
than the maximum halo assembly bias signals seen in simulations.

However, one caveat in this analysis is that using only high probability
members limits the use of the galaxy members which are closer to the cluster
center, due to the way the redMaPPer algorithm assigns membership
probabilities. It can be seen in numerical simulations that the assembly bias
signal reduces in strength if the concentration derived from subhalos at a
smaller radii is used to subdivide clusters at fixed halo mass in order to
measure the assembly bias signal. Thus it is unclear if the reduced assembly
bias signal seen in observations when using the new proxy is a result of such a
selection or only a result of a reduction in projection effects.

In the second study, \citet{BuschWhite2017} mimicked the optical
cluster finding algorithm in photometric imaging data on mock catalogs
to investigate whether the large clustering difference in the two
subsamples claimed in \citet{Miyatake2016} is consistent with
$\Lambda$CDM.  Their results showed that they could reproduce the
observed clustering difference in the two subsamples seen in observations. The
presence of projection effects in optical cluster catalogs was key to
establishing the size of the signal observed. Although the seemingly large
clustering difference could be reproduced in mock catalogs from $\Lambda$CDM,
it remains unclear if the methodology used by \citet{Miyatake2016} and
\citet{More2016} can be unambiguously used to detect halo assembly bias.

Addressing this question is the the central motivation of this paper.
In order to carry out this investigation, we mimic the redMaPPer
cluster finding algorithm closely and apply it to two different galaxy
catalogs, one catalog which has assembly bias signal and the other one
where the existing assembly bias has been artificially erased. If both
the catalogs show an assembly bias signal following the analysis
strategy adopted by \citet{Miyatake2016} and \citet{More2016}, that
would indicate that their methodology cannot be used to confirm or
deny the existence of the assembly bias signal.

This paper is organized in the following manner. In Section 2, we
describe the publicly available simulation data as well as the
observational data we use in our analysis. Section 3 presents the
simplified algorithm we use in order to mimic optical cluster finding
in photometric data, our estimation of the compactness of galaxy
clusters as a proxy of concentration of halos, steps to generate the
catalog without assembly bias, and computation of correlation
functions used in this paper. We presents our primary results in
Section 4 and conclude in Section 5.

\section{Data }

\subsection{Millennium Simulation: galaxy catalogs}

We use the Millennium Simulation (\citet{springel_etal05}), a
$2160^{3}$ particle cosmological N-body simulation of a flat
$\Lambda$CDM universe with box size of $500\mpch$ and mass
resolution of $8.6\times10^{8}\msunh$. The cosmological
parameters used in this simulation are the matter density parameter
$\Omega_{{\rm m}}=0.25$, the baryon density parameter $\Omega_{{\rm
b}}=0.045$, the spectral index of initial density fluctuation $n_{\rm
s}=1$, their amplitude specified by $\sigma_{8}=0.9$, and ${\rm
H_{0}=73{\rm km}{\rm s^{-1}}{\rm Mpc^{-1}}}$. Haloes and their
self-bound subhaloes were identified using the subfind algorithm
(\citet{springel2001}). The particular galaxy population used in this
paper was created using the semi-analytic model for galaxy formation
described in detail in \citet{guo_etal11b}. In this paper, we use the
same threshold as in \citet{BuschWhite2017} to select galaxies from a
snapshot at $z=0.24$. We restrict ourselves to galaxies with $i$-band
absolute magnitude $M_{i}<-20.14$ and with the specific star formation
rate (SSFR) below $1.5\times10^{-11}h{\rm yr}^{-1}$. The chosen
magnitude limit is very close to the one corresponding to the
redMaPPer luminosity limit of $0.2L_{\star}$ at $z=0.24$ (see
\citet{Rykoff2012}), and the cut in the SSFR of the galaxies define a
class of red galaxies.  This selection criterion gives us $925114$
galaxies. 

\subsection{redMaPPer galaxy clusters}

In order to compare our simulation based results to observations in
the real Universe, we will make use of the redMaPPer galaxy cluster
catalog. Using the imaging data processed in the Sloan Digital Sky
Survey Data release 8 (\citet{Aihara_etal2011}), \citet{Rykoff_etal2014}
ran the red-sequence cluster finder, redMaPPer (\citet{Rykoff_etal2014,
Rozo2014, Rozo2015, Rozo2015_2}), in order to find galaxy clusters. The
galaxy clusters identified by redMaPPer have been assigned richness as
well as redshifts. In order to mimic the selection criteria in
\citet{More2016}, we choose galaxy clusters with richness
$20\leq\lambda\leq100$ and redshifts between $0.1<z<0.33$. More than
$80\%$ of galaxy clusters in the catalog have a spectroscopic
redshift. We remove the remaining clusters which do not have a
spectroscopic redshift from our analysis.  We also use the random
catalogs provided along with the redMaPPer cluster catalog. These
catalogs contain corresponding position information, redshift,
richness as well as a weight for each random cluster.

\subsection{BOSS DR12 LOWZ sample}

We will carry out a cross-correlation of the redMaPPer galaxy clusters
with spectroscopic galaxies in order to study halo assembly bias. We
use the spectroscopic galaxies in the large scale structure catalogs
constructed from SDSS DR12 (\citet{Alam2015}). In particular we will use
the LOWZ sample, since it has a large overlap in redshift range as our
galaxy cluster sample. We restrict ourselves to LOWZ galaxies with
redshifts between $0.1<z<0.33$, the same redshift range as our galaxy
clusters.  The galaxy catalogs also come with associated random galaxy
catalogs that we use in order to perform our cross-correlation
analysis.

\section{Methods\label{sec:methods}}

The goal of this paper is to test the suitability of the methodology
adopted by \citet{Miyatake2016} and \citet{More2016} to detect the
assembly bias signal. As stated in the introduction,
\citet{BuschWhite2017} showed that the large clustering difference
observed by \citet{Miyatake2016} and \citet{More2016}  in their
cluster subsamples divided by $\avrmem$ can be reproduced by running
an analogue of the redMaPPer algorithm on mock galaxy catalogs.
However, it is not clear whether the methodology used by
\citet{Miyatake2016} and \citet{More2016} can unambiguously be used to
detect halo assembly bias. Therefore to test this, we create a galaxy catalog
where we erase the inherent assembly bias signal explicitly. We do
this by exchanging the cluster-centric positions of galaxies in halos
of the same mass. We call such galaxy catalogs ``shuffled''.  

We also implement an analogue of redMaPPer algorithm and apply it to
both the original (hereafter, non-shuffled) and the shuffled catalogs.
We will compare the difference of clustering amplitude from the
shuffled catalogs with that obtained from the non-shuffled catalogs.
In this section, we first describe our implementation of redMaPPer
algorithm and the shuffled catalogs in Sec. \ref{subsec:redMaPPer} and
\ref{subsec:shuffled} respectively, and then we discuss how we measure
assembly bias signals and splashback radius in Sec. \ref{subsec:corr}.

\subsection{redMaPPer algorithm for mock catalogs\label{subsec:redMaPPer}}

We implement a cluster identification algorithm based on the redMaPPer
algorithm (\citet{Rykoff_etal2014})using the projected distribution of
galaxies. The algorithm uses multi-wavelength imaging data and
identifies overdensities of red galaxies. The red galaxy spectrum
shows a drop in their flux downward of 4000 Angstrom. This
break moves through the different optical wavelength bands for
galaxies with varying redshifts. In the absence of spectroscopic
redshifts, this photometric feature helps to distinguish galaxies at
different redshifts occupying the same position in the sky.

Ideally the redMaPPer algorithm can be directly applied to mock galaxy
catalogs if the colors of galaxies are realistic. Unfortunately, redMaPPer 
has not been run on semi-analytical models due to either
the non availability of light cone data or due to difficulties in
reproducing the widths of the red sequence of galaxies in such models,
which is critical for the identification of clusters (E. Rozo, E. Rykoff, priv. comm.). 
We make the simplifying assumption that the photometric redshift filter 
used to group galaxies along the redshift direction will have a poor resolution
and therefore identify galaxies within a certain distance along the LOS as
cluster members. These projection effects are expected to cause a systematic
bias in the measurement of cluster observables. The effective LOS distance
within which we consider galaxies as members of the cluster, will be called the
projection length of the selection cylinder, or projection length in short.

We will vary the value of this projection length to assess its impact on our
conclusions as well as to compare with the actual data.

Our algorithm is an improved version of the simplistic implementation
of \citet{BuschWhite2017}. In particular our algorithm will also
assigns a membership probability $p_{{\rm mem}}$ to each galaxy in a
manner similar to redMaPPer. 

\subsubsection{Algorithm}

In redMaPPer, the probability that any galaxy found near a cluster center is
part of the cluster with richness $\lambda$ is denoted by $\pmem$ and is given
by
\begin{equation}
\pmem = \frac{\lambda u(x|\lambda)}{\lambda u(x|\lambda) + b(x)}\,,
\label{eq:pmem}
\end{equation}
where $x$ denotes properties of galaxies and includes the projected
cluster-centric distance of the galaxy, $u(x|\lambda)$ denotes the normalized
projected density profile of the galaxy cluster, and $b(x)$ denotes
the background contamination. The total richness of a galaxy cluster
should satisfy
\begin{equation}
\lambda = \sum_{R<R_{\rm c}(\lambda)} p_{\rm free} \pmem(x|\lambda)\,,
\label{eq:lambda}
\end{equation}
where the sum goes over all members of a galaxy cluster within a
cluster-centric radius $R<R_{\rm c}(\lambda)$ and the line-of-sight
separation $|\pi|<d_{\rm cyl}^{\rm max}$. The probability
$p_{\rm free}$ indicates the prior that the galaxy does not belong to
any other richer galaxy cluster. The radial cut scales with $\lambda$
such that
\begin{equation}
R_{\rm c}(\lambda) = R_0 (\lambda/100.0)^{\beta}\,,
\label{eq:radius}
\end{equation}
where $R_0=1.0\mpch$ and $\beta=0.2$ as adopted in redMaPPer. We will
use three different values of $d_{\rm cyl}^{\rm max}$ ranging from
$60 \mpch$ to $250 \mpch$ in our analysis.

The profile $u(x|\lambda)$ is the projected NFW profile \citep{NFW97,
Bartelmann96} and is truncated smoothly at a projected radius $R=R_{\rm c}$
with an error function as described in \citet{Rykoff_etal2014}. The background
$b(x)$ is assumed to be a constant to model the uncorrelated galaxies in the
foreground and the background.

At first, all the galaxies in the catalog are considered potential cluster
central galaxies and are given a probability $p_{\rm free}=1$ to be
part of any cluster. We start our percolation of galaxy clusters by
rank ordering galaxies by their stellar mass. We compute richness
$\lambda$ for each candidate central galaxy by taking all the galaxies
within a radius of $0.5\mpch$ and $|\pi| <
d_{\rm cyl}^{\rm max}$.  In this first step, the membership
probability for all galaxies is set to be unity if it is within the
above defined cylinder. The probability $p_{\rm free}$ of each galaxy
within this cylinder is then updated to be $p_{\rm free}=0$. The
percolation steps continue by going down the rank ordered list of
galaxy centers and ignoring those which have been assigned a $p_{\rm
free}=0$. The purpose of this first step is to find an overall
over-density regions which potentially have clusters. We eliminate all
the candidates with $\lambda<3$ from the list. 

Then, we reset $p_{\rm free}=1$ for all galaxies. We rank-order the
clusters in a descending order based on the preliminary richness
$\lambda$ and take percolation steps iteratively.  Note that we
simplify the rank-ordering by only using $\lambda$, while the actual
redMaPPer algorithm implemented in \citet{Rykoff_etal2014} rank
ordered cluster center candidates by $\lambda$ as well as the $r$-band
absolute magnitude of the galaxies. Starting from richest cluster, we
take the following steps.
\begin{enumerate}
\item Given the $i^{\rm th}$ cluster in the list, recompute $\lambda$
and the membership probability based on the hitherto percolated galaxy
catalog.
\item Determine the cluster center and clustering probability by
numerically solving for Eq.~\ref{eq:lambda}. Take all the galaxies
within the radius of $R_{c}(\lambda)$ and the projection length
$\Delta \pi$. If there is a brighter galaxy than the current
cluster center galaxy, consider the brightest one as a new center. 
\item Recompute $\lambda$ with respect to the new central galaxy based
on Eqs. \ref{eq:pmem} and \ref{eq:lambda}.
\item Update the probability $p_{\rm free}$ for each galaxy to be
$p_{\rm free}(1-\pmem)$ based on their membership probabilities of
the current cluster. If $p_{\rm free}>0.5$, then these galaxies are eliminated
from the list.
\item Repeat the steps for the next galaxy cluster in the ranked list.
\end{enumerate}
Note that we model the photometric redshift uncertainty by using
various projection lengths. To see whether the modeling of the
projection affects our results, we also tried some window functions
along the line-of-sight to model photometric redshift uncertainties in
a slightly different manner, but we did not see significant changes in our
results.

\subsubsection{Definition of Concentration $R_{{\rm mem}}$}

As a probe of cluster concentration measured from the distribution of member galaxies in clusters, \citet{Miyatake2016} and
\citet{More2016} use the mean projected length of membership
galaxies from the cluster center, $\avrmem$ defined as 
\begin{equation}
\avrmem=\frac{\Sigma p_{{\rm mem},i}R_{{\rm mem},i}}{\Sigma p_{{\rm mem},i}},
\end{equation}
where $p_{{\rm mem},i}$ and $R_{{\rm mem},i}$ are the membership
probability and the projected distance from the cluster center of
the $i$th member of galaxies in the cluster\footnote{The distribution of the true cluster satellites is expected to follow the concentration of dark matter within the halo (e.g., \citet{Han_etal2018}), even though subhalos detected in numerical simulations do not show this behaviour (e.g., \citet{More2016, vdBosch_etal2018a,vdBosch_etal2018b}).}. As concentration has
some dependencies on halo mass, $\avrmem$ depends on redshift and richness
$\lambda$, which is a proxy for cluster mass. In
order to measure the assembly bias signals, we need to split the
cluster sample into two subsamples with the same mass and which have
different $\avrmem$. To do this, we use 10 equally spaced bins both in
redshift and \textgreek{l} and obtain the median of $\avrmem$ as a
function of redshift and $\lambda$. We then refer to clusters with
$\avrmem<\overline{\avrmem}(z,\lambda)$ as
``small-$\avrmem$ clusters'' and clusters with $<R_{{\rm
mem}}>>\overline{\avrmem}(z,\lambda)$ as ``large-$<R_{{\rm
mem}}>$ clusters''. This ensures that the two cluster subsamples have
the same richness, and yet different internal structure.  In order to
mimic the methodology in \citet{Miyatake2016} and \citet{More2016}, we
use all the member galaxies including those with low membership
probabilities.

\subsection{Shuffled Catalogs \label{subsec:shuffled}}

\begin{figure}
\includegraphics[width=0.45\textwidth]{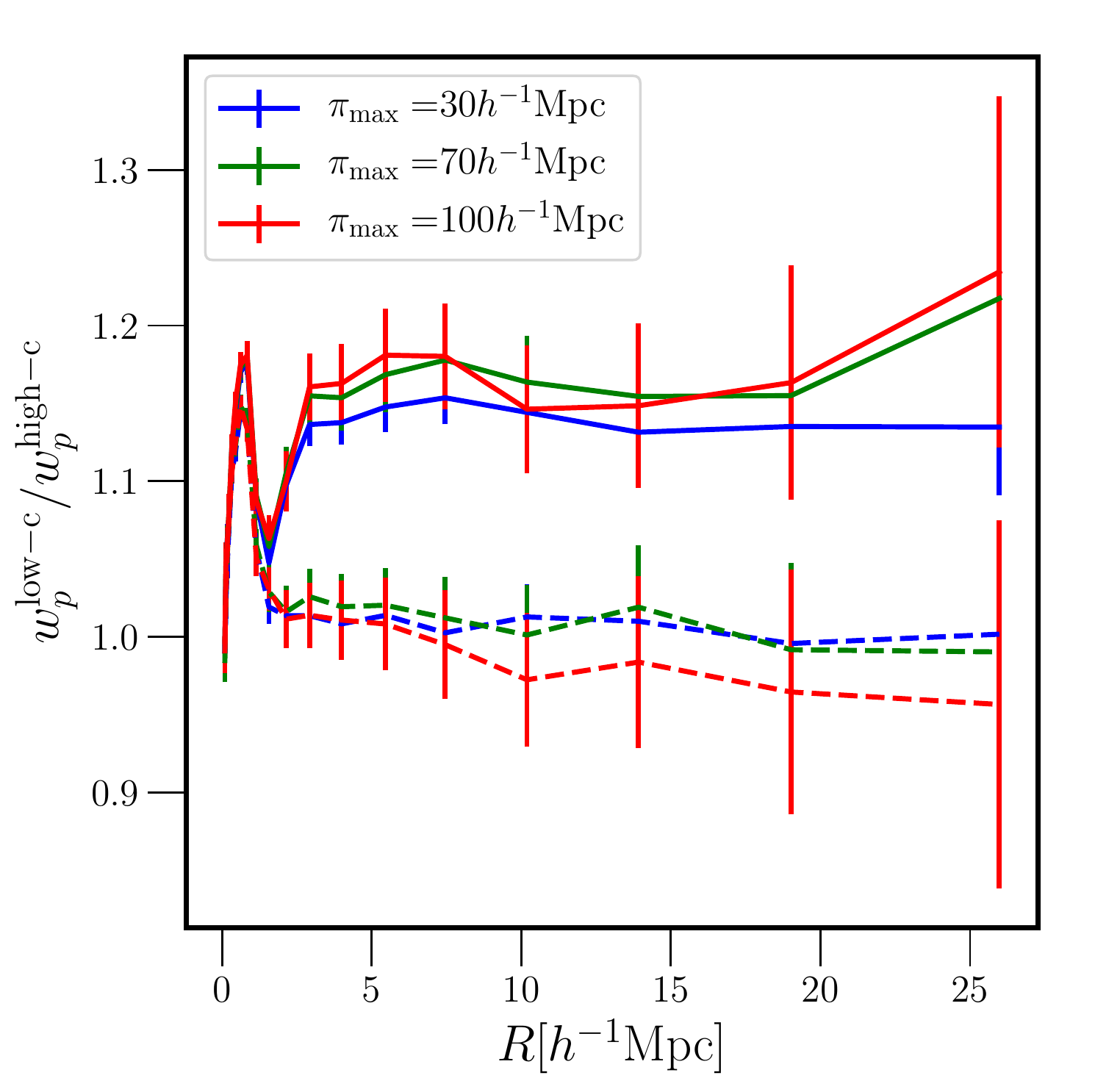}

\caption{\label{fig:assembly_concen} The figure shows our method of erasing
assembly bias explicitly by using shuffled catalogs. Ratio of cluster-subhalo
cross correlation functions between galaxy clusters with high and low dark
matter concentration from the MultiDark Planck simulation at $z=0.25$. Each
line corresponds to different integration scale $\pi_{\rm max}$. The solid
lines correspond to the ratios measured from the non-shuffled catalogs, while
the dashed lines are from the shuffled catalogs. Since the secondary parameter
dependence is erased in the shuffled catalogs, we do not see any assembly
biases from the shuffled catalogs.} \end{figure}



Halo bias primarily depends on halo mass. However, halo bias depends
on additional properties beyond their mass due to different assembly
history of halos. This additional dependence of halo bias is called
assembly bias. Since assembly bias is a secondary dependence of halo
bias, we can erase the assembly bias from the catalog by shuffling the
positions of the same mass halos. To do so, we first rank order halos
by mass and split them into a series of bins with constant halo mass
range. We use 10 bins, but the choice of the number of bins is arbitrary. For
each mass bin, we shuffle the central positions of the host halos while
maintaining the positions of the subhalos within their parent halos.


For demonstration purposes, we use halo catalogs from Multi-Dark Planck (MDPL)
simulation (\citet{Klypin_etal2016}) and select cluster-sized halos (i.e.,
$M_{\rm halo}>10^{14}h^{-1}{\rm M}_{\odot}$) splitting into low/high-concentration subsamples, where
concentration is measured from dark matter profiles.
Fig.~\ref{fig:assembly_concen} shows the ratio of projected correlation
functions of low-concentration halos divided by those of high-concentration
samples for several different integral scale along the line-of-sight.
The catalogs with assembly bias (i.e., non-shuffled catalogs) exhibit
the expected size of assembly bias signal for cluster-sized halos,
while the subsamples from the shuffled catalogs do not show any bias
differences. We choose several different maximum projection 
length $R=10\mpch$ , $30\mpch$, and $100\mpch$ to compute
projected correlation functions defined in Eq. \ref{eq:corr_fn-1}.
Bias ratios from both catalogs show little dependence
on the integral scales.


Fig.~\ref{fig:assembly_concen} shows that the shuffled catalogs do not 
exhibit the assembly bias signals as expected. If the projected correlation functions 
of the redMaPPer clusters using the shuffled and the non-shuffled catalogs 
exhibit the similar result as Fig.~\ref{fig:assembly_concen}, it
implies that the assembly bias signals detected by
\citet{Miyatake2016} and \citet{More2016} is actually a manifestation
of assembly bias.

\subsection{Correlation functions \label{subsec:corr}}

We compute the projected cross-correlation functions for each cluster
sample and the galaxy sample as 
\begin{equation}
w_{p}(R)=\int_{0}^{\pi_{\rm max}}dr_{\pi}\xi_{gc}(R,r_{\pi}),\label{eq:corr_fn-1}
\end{equation}
where $\pi_{\rm max}$ is the maximum integral scale, $\xi_{gc}$ is a
two-dimensional cross correlation functions between clusters and
galaxies. We use several different integral scales $\pi_{\rm max}$ to
evaluate the contamination from the foreground/background galaxies.
\citet{BuschWhite2017} and \citet{Zu2017} argued that the proxy of the
concentration $\avrmem$ is strongly affected by projection effects and
therefore the parameter is correlated with the surrounding density
field. We explicitly evaluate the level of contamination due to the
projection effects by varying the integral scale $\pi_{\rm max}$. We
use 102 jackknife regions in order to compute the error in the
measurements of $w_{p}(R,r_{\pi}<\pi_{\rm max})$.  The typical size of
each of these jackknife patches is about $10\times10$ square degrees,
which corresponds to roughly $100\times100(\mpch)^{2}$ for
our cluster and galaxy samples. If the signal of assembly bias is
purely due to foreground/background galaxies, we expect to see a
dependence on the size of signal when varying $\pi_{\rm max}$.

\subsection{Comparison with Observational Data}

\begin{figure}
\includegraphics[width=0.45\textwidth]{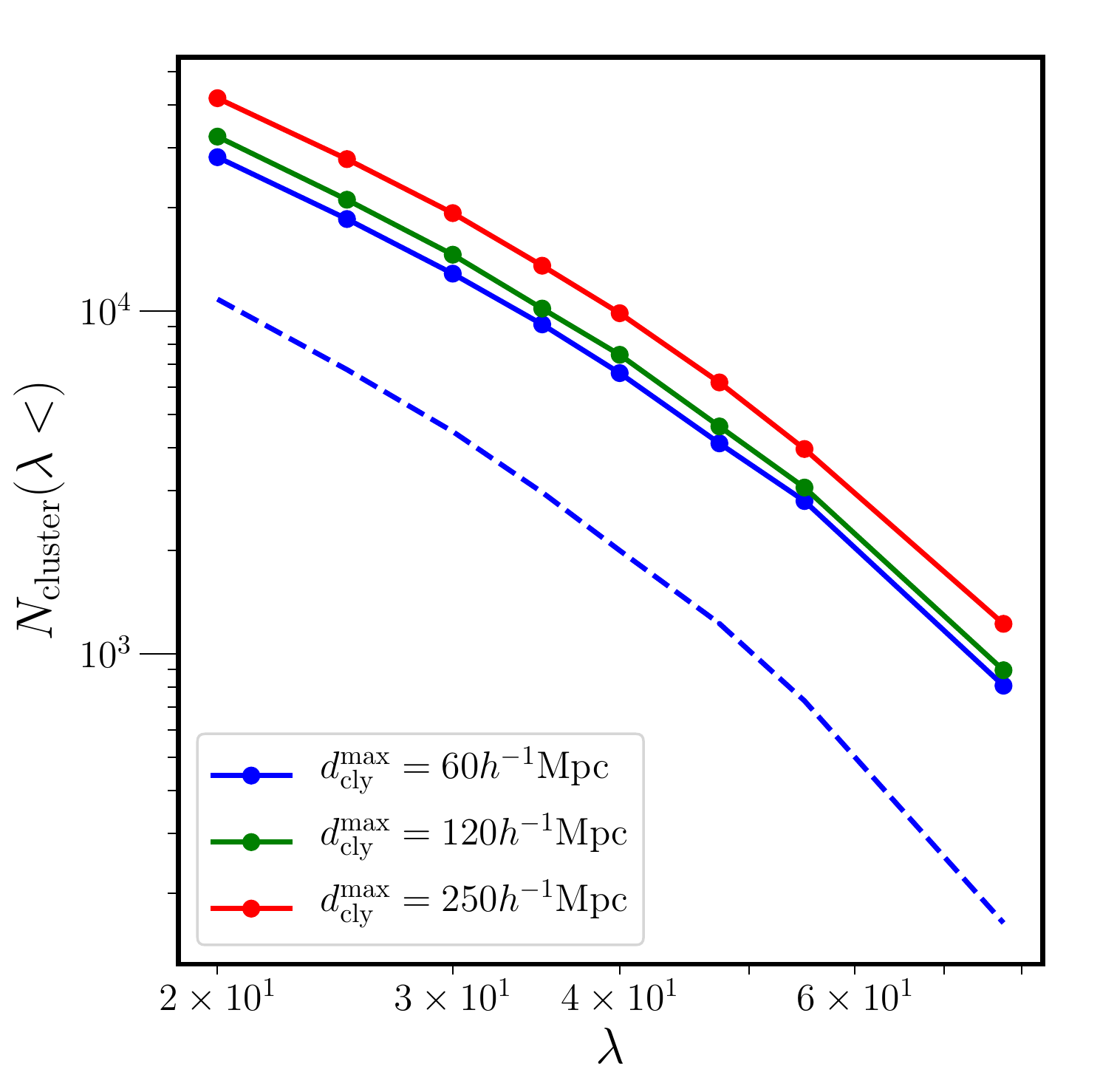}

\caption{\label{num_clusters}The number of clusters as a function of richness
$\lambda$. Different colors correspond to the projection used to
identify clusters. The dashed line corresponds to the number of clusters
identified by redMaPPer from DR8.}
\end{figure}

\begin{figure*}
\includegraphics[width=0.4\textwidth]{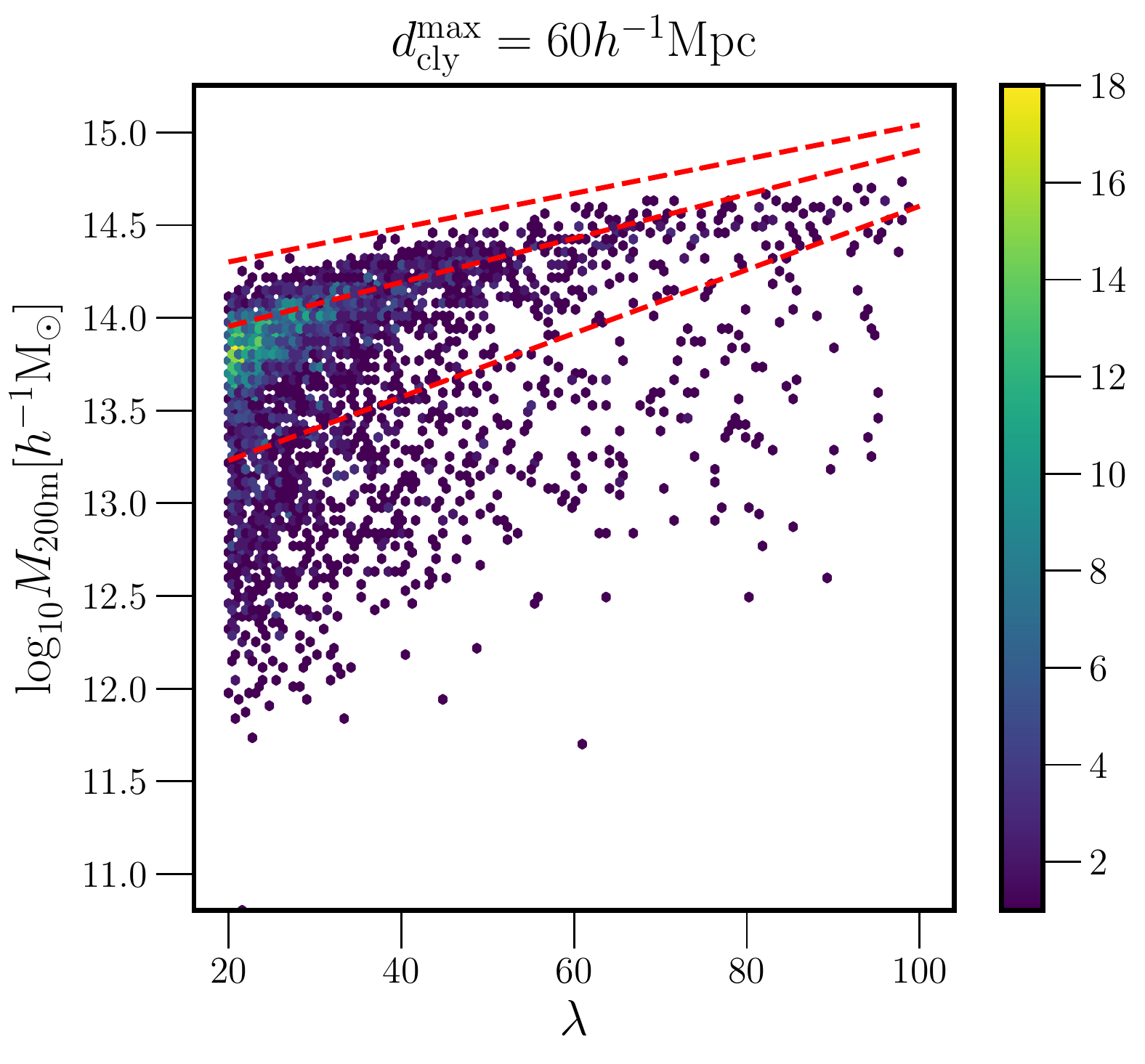}\includegraphics[width=0.4\textwidth]{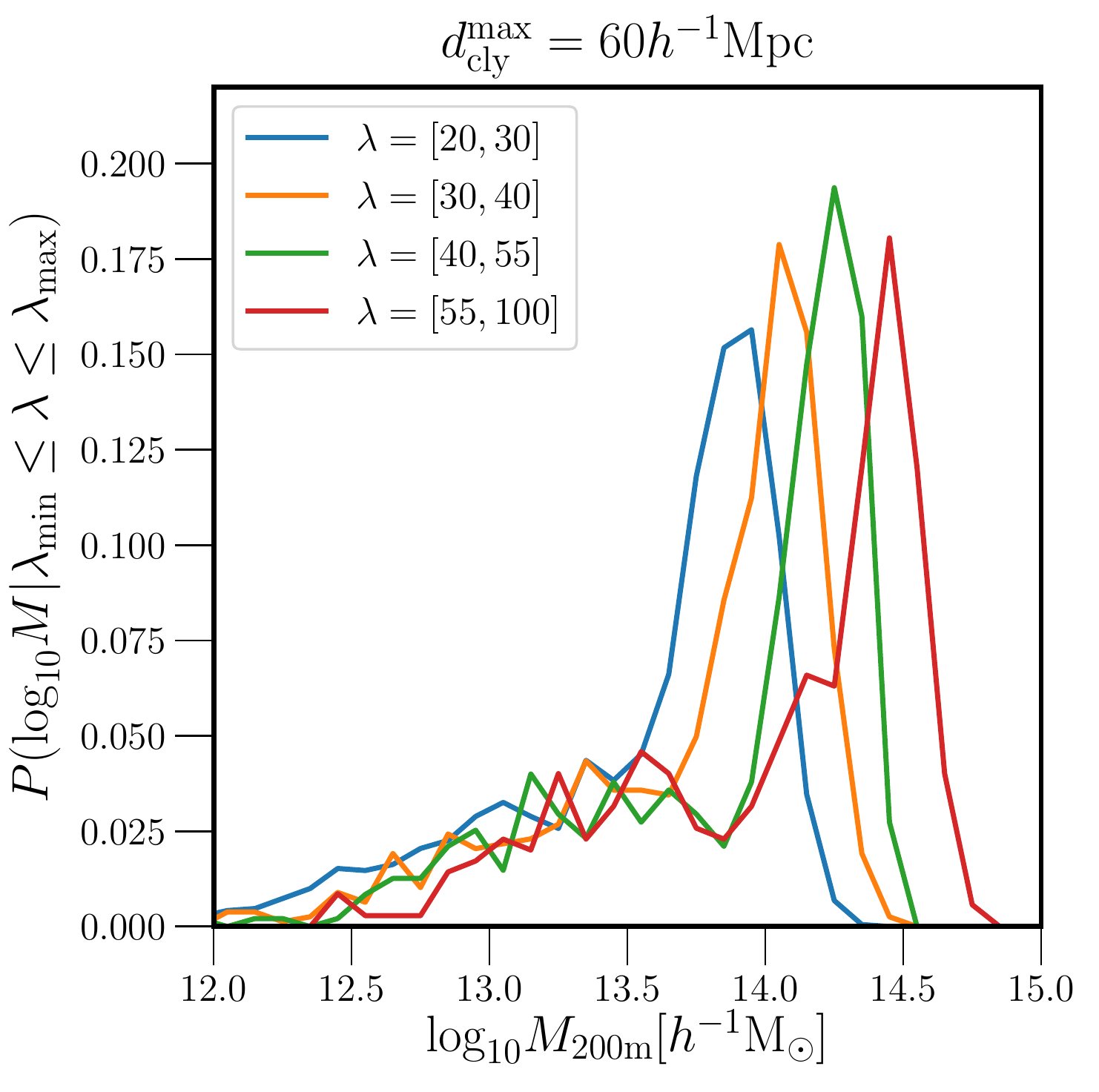}

\caption{\label{fig:mass-richness}(Left) Comparison of galaxy clusters found
using our mock redMaPPer method compared to inferences from real redMaPPer
clusters. The left hand panel shows the halo mass-richness relation of the
clusters selected using $d_{\rm cyl}^{\rm max}=60\mpch$. The red dashed lines
are the mean and standard deviation from \citet{Murata_etal2018}. The right
hand panel shows the histograms of halo mass for different richness bins.}
\end{figure*}

In this section, we compare properties of clusters identified by our
analogue of redMaPPer and weak lensing analysis using redMaPPer
clusters by \citet{Murata_etal2018} to make sure that our version of
redMaPPer can produce salient features of the observations. In Fig.
\ref{num_clusters}, we plot a cumulative number of clusters in $(1h^{-1}{\rm Gpc})^3$ as a
function of $\lambda$.  

The solid lines with different colors show the result of running our
mock redMaPPer algorithm on the Millennium galaxy catalog with
different projection lengths to account for the uncertainties along
the line-of-sight, while the dashed line shows the number of redMaPPer
clusters in the original DR12 galaxy catalog. More number of clusters are 
identified by using longer projection lengths.  This is because the
number of member galaxies in the cylinder increases as the projection
lengths and even slightly over-dense region can be identified as a
cluster.  This trend was also noted by \citet{BuschWhite2017}.  The figure also shows that
our implementation of redMaPPer can reproduce the relative trend in the cluster number
consistent with observations. Note that our comparison with the
observations here can only be qualitative rather than quantitative.
This partially stems from the fact that cosmological parameters used in the Millennium
simulations are different from the Planck cosmology. In particular, due to the large
value of $\sigma_{8}$ used in the Millennium simulations, the number
of clusters can be about $25\%$  larger than that
for the Planck cosmology at fixed halo mass (\citet{BuschWhite2017}). In
addition, any differences between the mapping between colors of galaxies and
their star formation rate cuts that we have used on the Millennium galaxy
catalog to select our members, can cause systematic differences in richness and
result in a larger number of galaxy clusters at a fixed richness. It is likely that our input galaxy catalog
contains more number of galaxies, which causes the increase in the number of clusters 
and the decrease in the mean halo masses (see Fig. \ref{fig:mass-richness} )  at the fixed richness. 

Next, we compare the richness-mass relation of the clusters identified
from the Millennium simulation and the constraints from the weak
lensing analysis of redMaPPer clusters carried out by
\citet{Murata_etal2018}. The left panel of Fig.
\ref{fig:mass-richness} shows the overall mass-richness relation of
the clusters. The red dashed lines show the median and the 16th and
84th percentiles of the mass distribution for the redMaPPer clusters
with $20\leq\lambda\leq100$ as found by \citet{Murata_etal2018}. The
scatter plot shows the halo mass richness relation of clusters
identified in our mock run on the Millennium simulation. Our optical
clusters also capture the same features as the observed mass-richness
relation. The right panel shows the distribution of halo mass as a
function of richness, which is also consistent with
\citet{Murata_etal2018} including the extended low-mass tail observed
at fixed richness. This low-mass tail is due to the projection effects that the
clusters contain some foreground/background galaxies and therefore
these low-mass halos gain large enough richness.

\section{Results}

In this section, we present our findings regarding the use of optically
selected clusters for measuring halo assembly bias as well as the splashback
radius. We reiterate that our goals are two-fold: can we use the methodology
adopted by \citet{Miyatake2016} and \citet{More2016} to (i) detect halo
assembly bias (ii) or obtain an unbiased measurement of the splashback radius
of optically selected galaxy clusters?

\subsection{Halo assembly bias\label{subsec:assemblyBias_sim}}

\begin{figure*}
\includegraphics[width=0.3\textwidth]{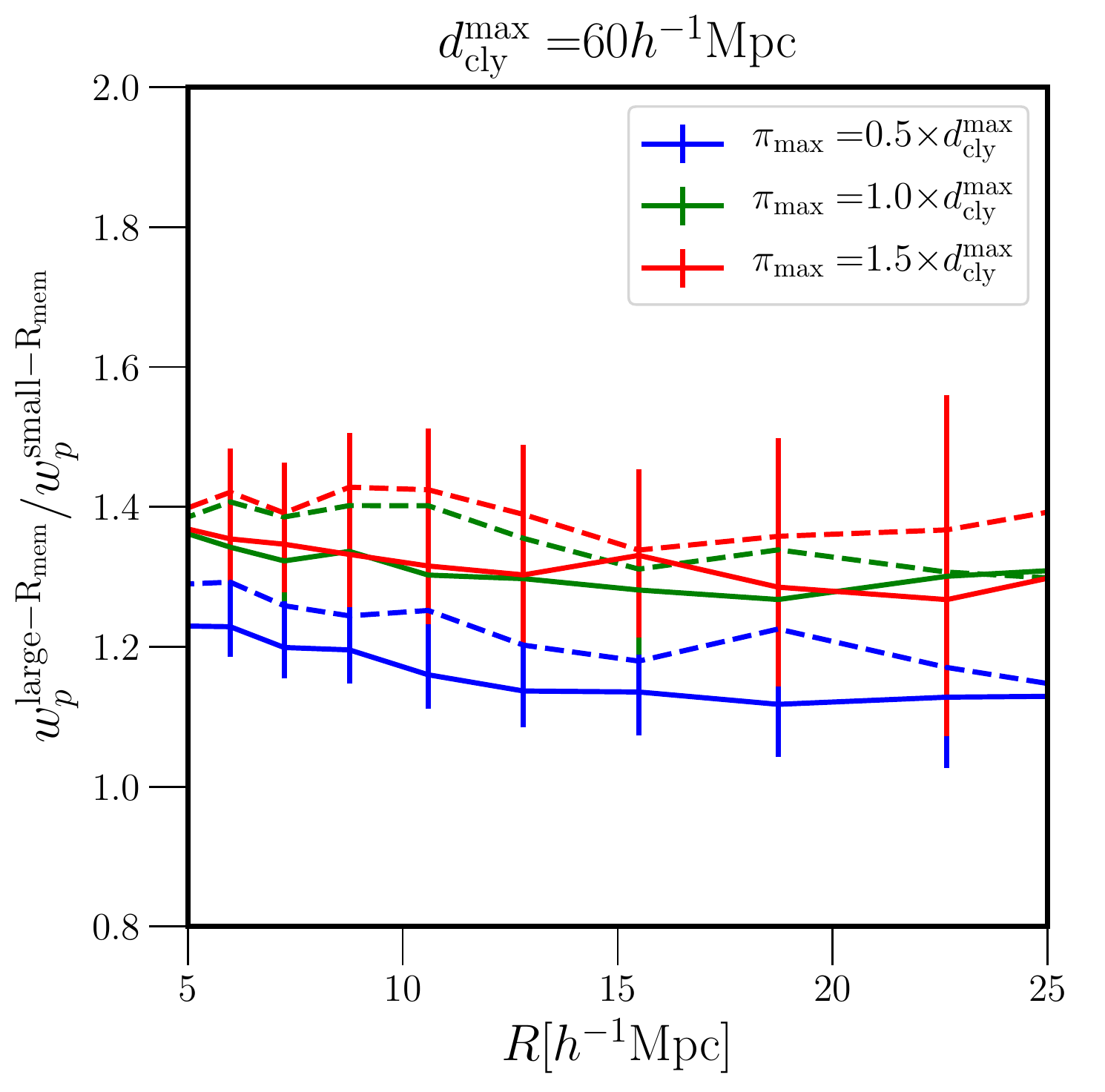}\includegraphics[width=0.3\textwidth]{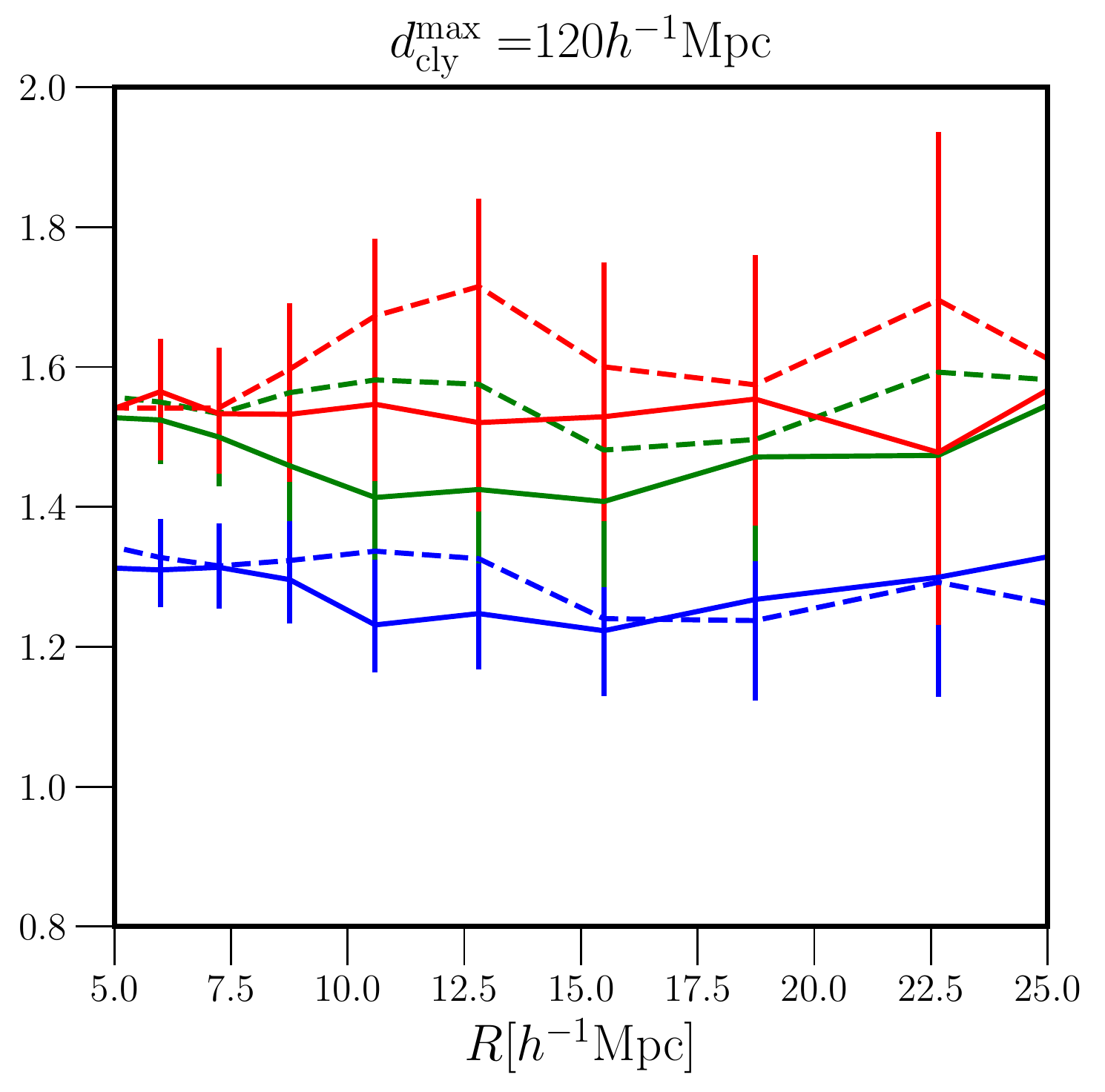}\includegraphics[width=0.3\textwidth]{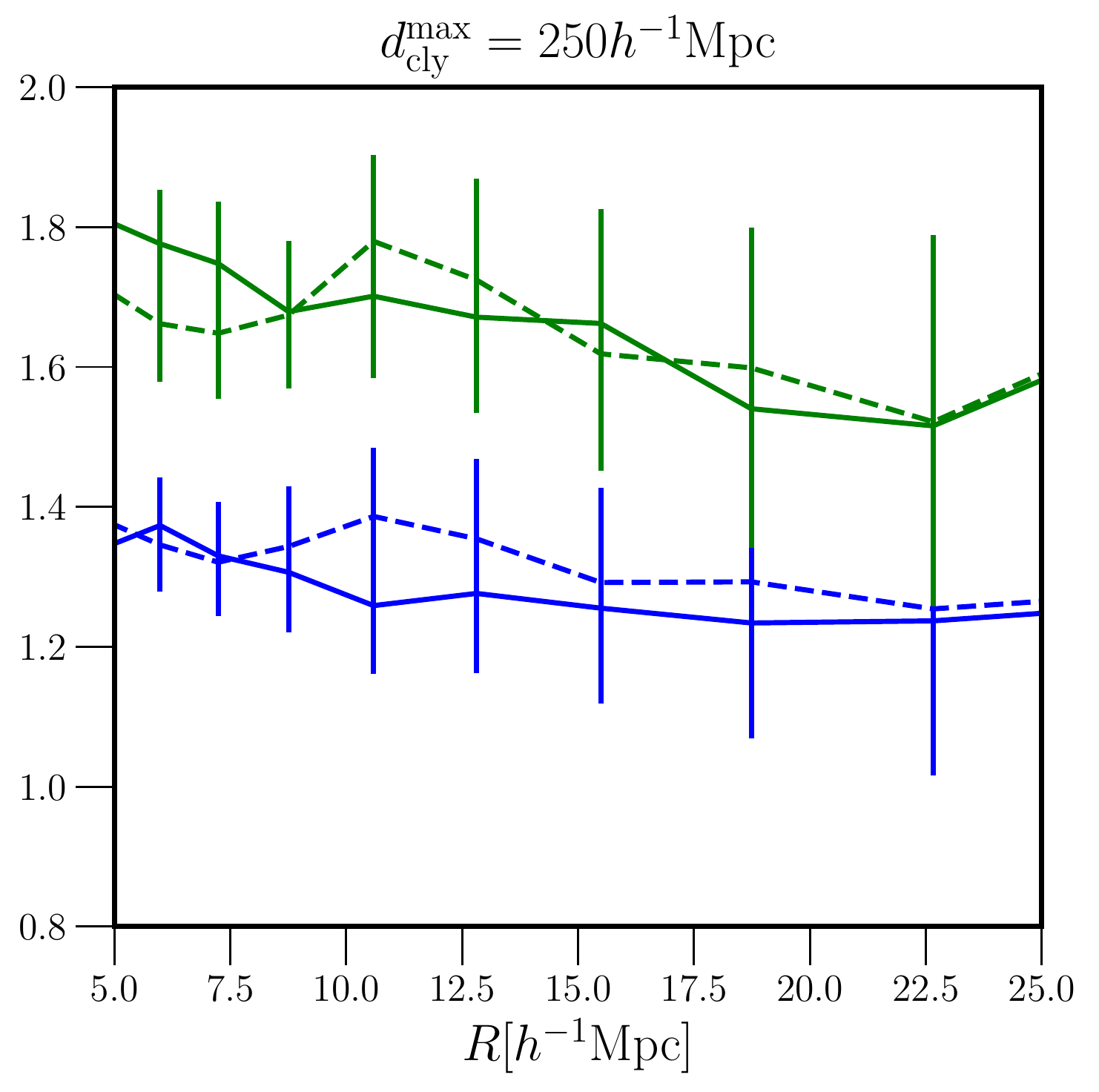}

\caption{\label{fig:assembly_s} Dependence of the ratio of the projected
correlation functions of the large/small-$<R_{\rm mem}>$ subsamples with
different integral scales $\pi_{\rm max}$. From left to right panels correspond
to the cluster samples with the projection lengths $d_{\rm cyl}^{\rm max}$ of
$60$, $120$, and $250\mpch$. Note that the projected correlation functions are
computed in redshift-space.}

\end{figure*}

We first present the results for halo assembly bias. We apply the algorithm
described in Section \ref{subsec:redMaPPer} to the galaxy catalog from the
Millennium simulation to select analogs of optically selected galaxy clusters.
The one free parameter in our algorithm is the effective cylinder length
$d_{\rm cyl}^{\rm max}$, galaxies within which get clubbed together in a
single observationally identified galaxy clusters. Therefore we will show
results for three different effective cylinder lengths, $d_{\rm
cly}^{\rm max}=60\mpch$, $120\mpch$, and $250\mpch$.  The
three panels of Fig.~\ref{fig:assembly_s} show the ratio of the projected
cross-correlation functions between galaxies and the subsamples of optically
selected galaxy clusters identified by the three different effective cylinder
lengths. We split the cluster samples into the large/small-$<R_{\rm mem}>$
subsamples, where  $<R_{\rm mem}>$ is a proxy of concentration used in
\citet{Miyatake2016} and \citet{More2016}.

In each panel, the solid lines correspond to the ratio of the
cross-correlation functions between the large/small-$<R_{\rm mem}>$
cluster subsamples from the non-shuffled catalogs.  Unlike
\citet{Miyatake2016}, we have integrated the correlation function out
to various integral scales, $\pi_{\rm max}=0.5\times d_{\rm cyl}^{\rm
max}$,  $d_{\rm cyl}^{\rm max}$, and $1.5\times d_{\rm cyl}^{\rm
max}$. Note that for the case of $d_{\rm cyl}^{\rm max}=250\mpch$, we
cannot integrate out to $1.5\times d_{\rm cyl}^{\rm max}$ because the
box size of the Millennium simulation is $500\mpch$. We see that there
is an apparent halo assembly bias signal seen in all three panels,
where the small-$\avrmem$ samples of galaxy clusters have a higher
clustering signal on large scales compared to the large-$\avrmem$
sample. The qualitative behaviour is very similar to that found by
\cite{Miyatake2016} and \citet{More2016}.


To ascertain that the difference in the clustering of the galaxy
clusters is indeed a result of halo assembly bias, we now rerun the
same measurements, but now by selecting clusters from a galaxy catalog
where halo assembly bias was intentionally erased (i.e., shuffled
catalogs). If the difference in clustering that we see is indeed due
to halo assembly bias, then the difference in the clustering signal
should vanish in this null test as with Fig.
\ref{fig:assembly_concen}. The resulting cross-correlation functions
of the clusters and galaxies in the shuffled galaxy catalogs are shown
as dashed lines in Fig.~\ref{fig:assembly_s}. We observe that within
the error bars of our measurement, we see consistent levels of
differences in the clustering amplitudes of the two subsamples,
between both the shuffled and the non-shuffled catalogs. The
conclusion remains the same even when we change the integral scales to
be smaller or larger than the projection length $d_{\rm cyl}^{\rm max}$. Given
that the same difference in clustering amplitude is also seen in the
shuffled catalogs, implies that the methodology used by
\citet{Miyatake2016} or \citet{More2016} cannot ascertain the
existence or absence of halo assembly bias.

In addition to little difference in the results from the shuffled
and non-shuffled catalogs, we see a strong dependence of the difference in the
clustering signal as a function of $\pi_{\rm max}$, at least for $\pi_{\rm
max}<=d_{\rm cyl}^{\rm max}$, while the signal stabilizes once $\pi_{\rm max}>d_{\rm
cyl}^{\rm max}$. The same behaviour is not seen in Fig.~\ref{fig:assembly_concen}, where
we use dark matter concentration to split cluster-sized halos into two subsamples. This
dependence on the integral scales can be possibly the evidence of projection
effects.

\begin{figure*}
\includegraphics[width=0.4\textwidth]{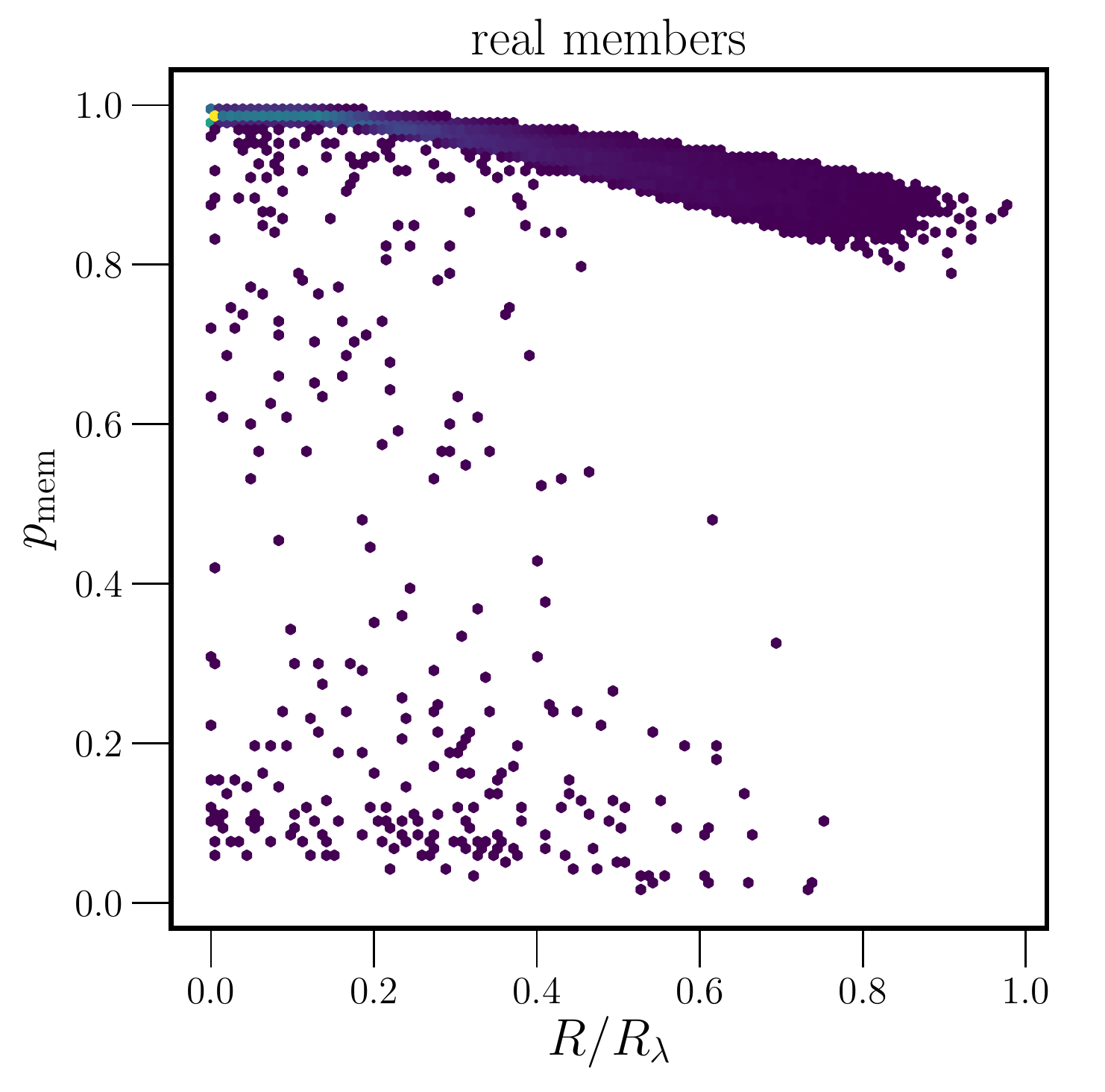}\includegraphics[width=0.4\textwidth]{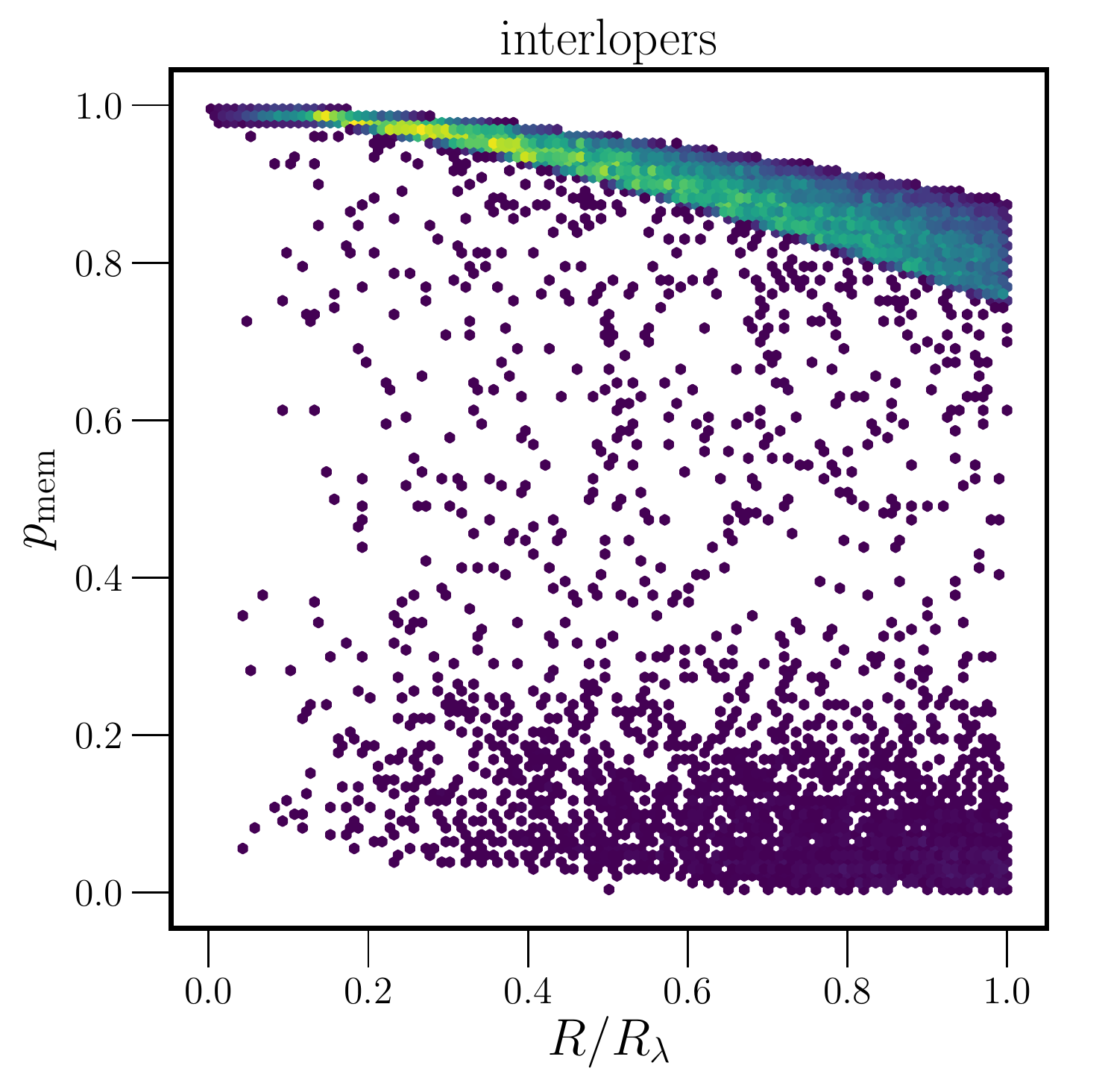}
\caption{\label{fig:pmemDist}The distribution of the cluster membership
probability $p_{\rm mem}$ as a function of the distance from the center of the
cluster normalized by the richness radius $R_{\lambda}$. The left panel shows
the distribution for the true members of the clusters, while the right panel
shows the distribution for the interlopers. These figures imply that we cannot
distinguish true members from interlopers based on the membership
probabilities. Note that we use the clusters identified by using the projection
length $d_{\rm cyl}^{\rm max}=60\mpch$.}

\end{figure*}

Next we explore the reason why the methodology does not work. As the
value of $d_{\rm cyl}^{\rm max}$ is much larger than the typical
radius of the halo, the optical members of the galaxy cluster
invariably include a number of members which are beyond the halo, but
lie along the line-of-sight to the primary halo.  In Fig.
\ref{fig:pmemDist}, we show the membership probability assigned by our
cluster finder as a function of the separation $R$ from the cluster
center, for real members as well as interlopers. Although the
membership probability does decrease as a function of radius, it can
be seen that the membership probability cannot really distinguish
between the interlopers and real members.  Thus the method of
separating the members from the interlopers using membership
probability as suggested by \citep{Zu2017}, does not really solve the
underlying problem.

\begin{figure}
\includegraphics[width=0.4\textwidth]{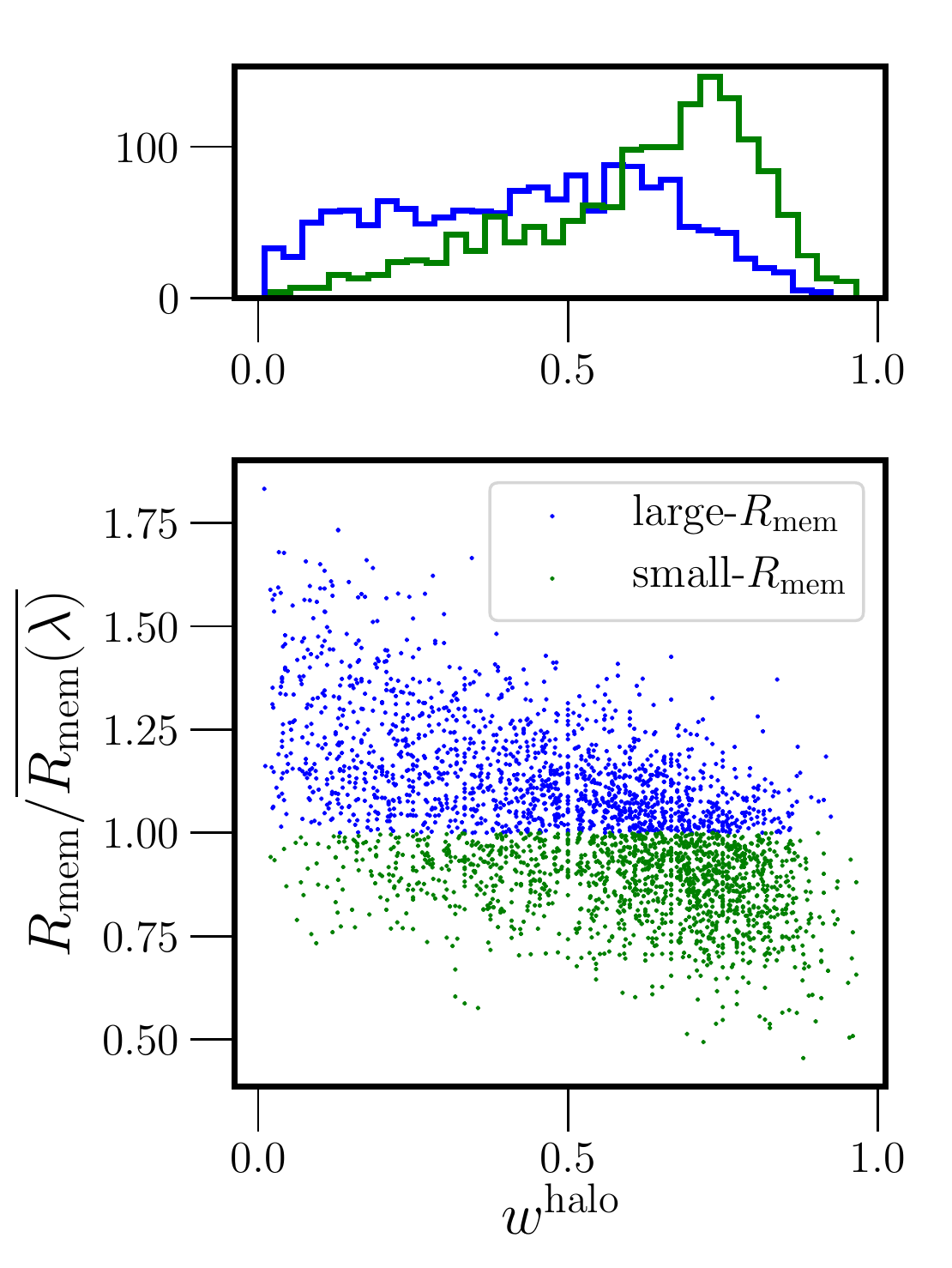}
\caption{\label{fig:weightDist}The figure shows the correlation between the
fraction of true halo members and $R_{\rm mem}/\overline{R_{\rm
mem}(\lambda)}$.  The clusters split into large/small-$<R_{\rm mem}>$
subsamples are shown in blue/green dots and are divided by the horizontal line
at unity. The large-$<R_{\rm mem}>$ clusters tend to contain larger fractions
of interlopers than the small-$<R_{\rm mem}>$ clusters. Note that we use the
cluster samples identified with the projection length of $60h^{-1}{\rm Mpc}$
here.}

\end{figure}

Due to this, the resultant value of $\avrmem$ from the assigned
members invariably consists of contributions from both the real
members as well as the interlopers. The division by $\avrmem$
automatically also results in a division based on the amount of
correlated large scale structure, with large $\avrmem$ corresponding
to larger values of the correlated structure. This then gets reflected
in the associated cross-correlation on large scales at fixed observed
richness.

Mathematically, we can write down
\begin{equation}
\avrmem = w^{\rm halo} \avrmem^{\rm halo} + w^{\rm int} \avrmem^{\rm int} \,,
\end{equation}
where $\avrmem^{\rm halo}$ and $\avrmem^{\rm int}$ denote the average
distance of the true members and the interlopers from the centers of
optically identified cluster centers. The weights $w^{\rm halo}$ and
$w^{\rm int}$ denote the weights which correspond to the fractional
contribution of the members and interlopers to the total richness of
the optically identified galaxy cluster system. Fig.
\ref{fig:weightDist} shows the distribution of the weight $w^{\rm
halo}$  for the case of the projection length
$d_{\rm cly}^{\rm max}=60\mpch$. The small-$<R_{\rm mem}>$
clusters tend to have larger real member fraction $w^{\rm halo}$
peaked around 0.8, while the distribution of $w^{\rm halo}$ for
large-$<R_{\rm mem}>$ clusters is more broadly distributed to
smaller values. This implies that the large-$<R_{\rm mem}>$
clusters are more correlated with large scale structure by including
more number of foreground/background galaxies.

The method adopted in \citet{Miyatake2016} and \citet{More2016} is marred by
the correlation coefficient between $\avrmem$ and the large scale overdensity.
The interpretation of the difference in clustering amplitude for different
$\avrmem$ clusters seen by \citet{Miyatake2016} and \citet{More2016}, thus
depends upon the relative strengths of these correlations with large-scale
overdensity as well as the weights $w^{\rm halo}$.  The use of the shuffled
galaxy catalogs completely erases out the correlation between the mean distance
of true members $<R_{\rm mem}>^{\rm halo}$ and the overdensity, thus making the
test entirely sensitive to just the projection effects. The fact that the
shuffled and the non-shuffled catalogs give similar differences in the
clustering amplitudes for the two cluster subsamples, implies that the results
are dominated by projection effects, and within the error bars, it would be
hard to disentangle the two.

\subsection{Implications for the measurement of the splashback radius}

\begin{figure}
\includegraphics[width=0.45\textwidth]{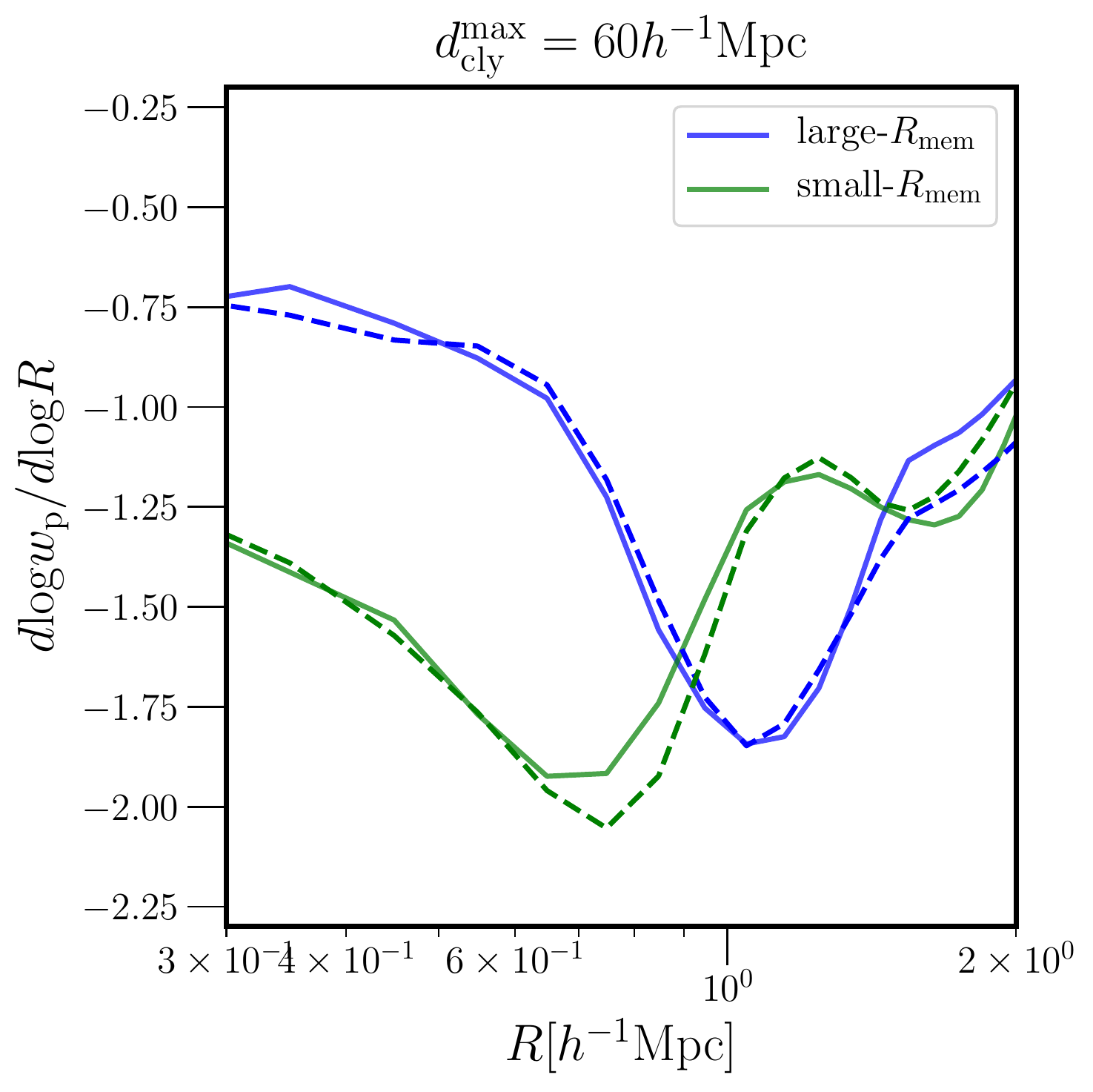}

\caption{\label{fig:splash}Logarithmic derivative profiles of the projected
correlation functions of clusters in the large-$<R_{\rm mem}>$ (blue)
and small-$<R_{\rm mem}>$ (green) subsamples using the projection 
length $d_{\rm cyl}^{\rm max}=60\mpch$.
 The integral scale $R$ of the correlation function is $100\mpch$. 
 Solid lines are the results from the non-shuffled catalogs,
while dashed lines are from the shuffled catalogs. Note that the location
of the splashback radius does not depend on the choice of the integral
scales. Note that these profiles are obtained by using the Savitzky-Golay
algorithm to smooth the measurements. We fit a third-order polynomial
over a window of five neighboring points.}

\end{figure}

Next, we measure the location of splashback radius for the optically selected
galaxy clusters from our simulations. The splashback radius corresponds to the
apocenter of the outermost shell in theoretical models of spherical symmetric
collapse \citep{adhikari2014splashback, shi2016outer}. These can be identified
by locating jumps in the density distribution \citep{ mansfield2017splashback},
or the average location of the apocenter of particles falling into the cluster
potential \citet{diemer2017splashback}. All these definitions follow the
general trend of a decreasing splashback radius with increasing rate of
accretion \citep{diemer2017splashback}.  In this paper, we adopt the convention
of \citet{MDK2015}, who define the location of the splashback radius
to be the location of the steepest logarithmic slope of the radially averaged
density profile, due to its ease of accessibility in data. For halos on galaxy
cluster scales, which on average are accreting matter at relatively faster
rates, this location corresponds to the location of the splashback radius.

We take the logarithmic derivative of the projected correlation functions and
identify the location of the steepest slope as the 2-dimensional splashback
radius \citep{DiemerKravtsov2014, MDK2015}. We obtain these logarithmic slopes
by using the Savitzky-Golay algorithm, to smooth the projected correlation
functions \citep{MoreASLcode}. Specifically we fit a third-order polynomial
over a window of five neighboring points and then use a cubic spline to
interpolate between these smoothed measurements. 

The results are shown in Fig.~\ref{fig:splash} for both the
non-shuffled and the shuffled cases. The solid lines are the results
for large-$\avrmem$ cluster subsamples, and the dashed lines for
small-$\avrmem$ subsamples. As is clear from the figures, the
large-$\avrmem$ subsamples (corresponding to low concentration
subsamples) have a larger splashback radius for both non-shuffled and
shuffled catalogs. We expect little difference in the results from non-shuffled and 
shuffled catalogs, because the distribution of galaxies within clusters (which
determines the location of splashback radius) does not change much even after shuffling.

Furthermore, the location of splashback radius does
not depend on either the choice of the projection length, or of the
scale of integration along the line-of-sight. Therefore we  decided
not to show in Fig.~\ref{fig:splash}. The locations of the steepest
slope of the two-dimensional cross-correlations in the different
$\avrmem$ samples seem to be robust against these various systematics.

However, we need to investigate if projection effects in optical
cluster finding can bias the measurements of the inferred location of
the 3-d splashback radius significantly. After all, it is the 3-d
splashback radius which contains information about the accretion rates
of galaxy clusters. Typically in studies that claim the detection of
the splashback radius, the projected cross-correlation between galaxy
clusters and galaxies is modeled under the assumption of spherical
symmetry to infer the location of the 3-d splashback radius. Given the
asymmetry introduced by the projection effects along the line-of-sight
due to optical cluster finding, it is important to gauge the impact of
the inaccuracy of this assumption.

\begin{figure}
\includegraphics[width=0.5\textwidth]{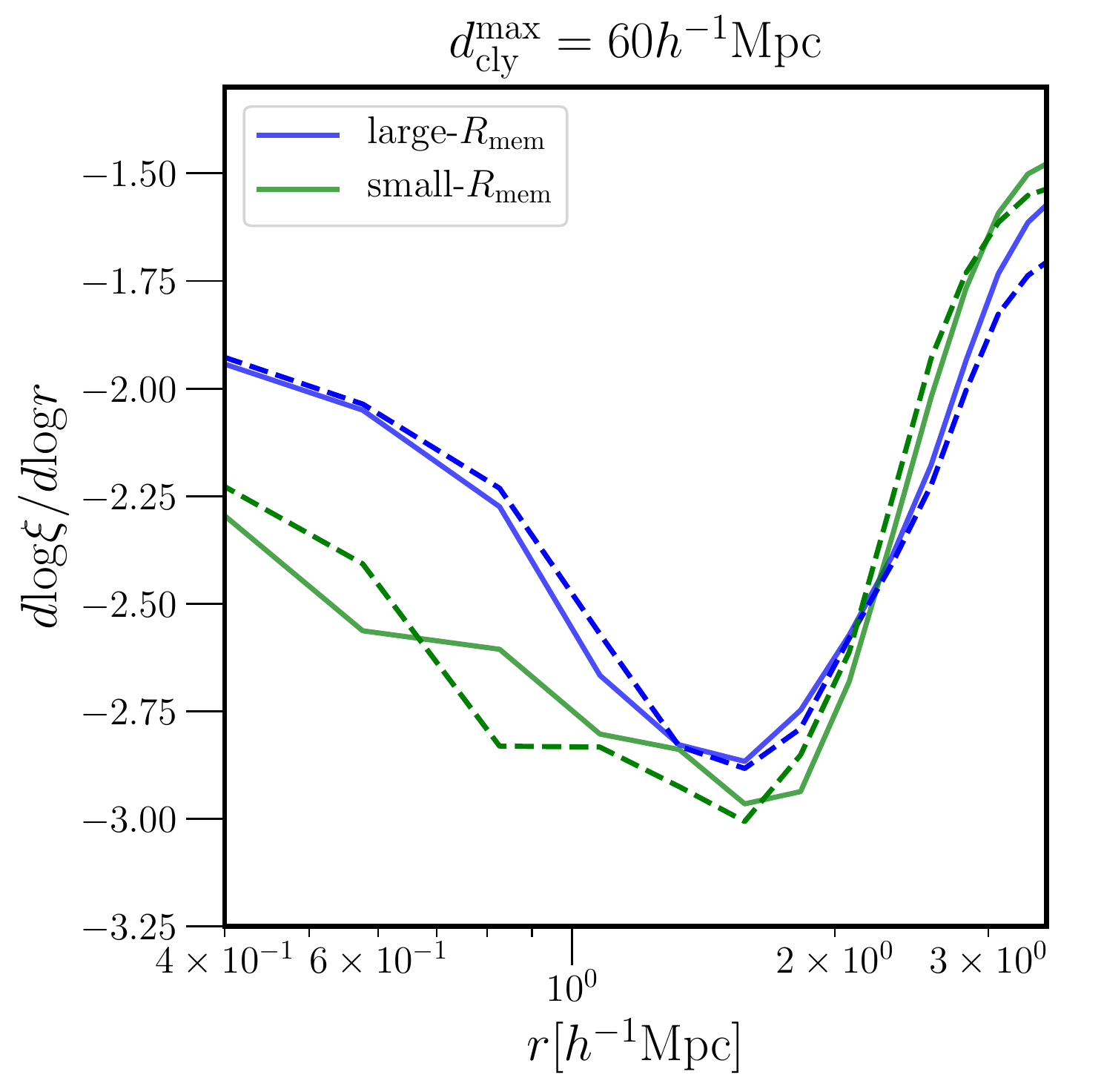}
\caption{\label{fig:splash_3D}Logarithmic derivative profiles of the three-dimensional
correlation functions $\xi_{\rm cg}(r)$ in real-space. The solid lines correspond to 
the derivative profiles from the non-shuffled catalogs, while the dashed lines are from
the shuffled catalogs. The splashback radii are larger than the
ones from the projected correlation functions and the locations are almost the same between
large/small-$<R_{\rm mem}>$ subsamples. 
}
\end{figure}

To test this, we compute the three-dimensional cross-correlation
functions $\xi_{\rm cg}(r)$ between galaxies and the
large/small-$\avrmem$ subsamples in real-space, as shown in Fig.
\ref{fig:splash_3D}. The splashback radii measured from the
three-dimensional correlation function are larger than the one from
the projected correlation function, and are almost the same unlike the
case of the projected ones. The similar location of the 3d splashback
radius is most probably a consequence of the similar halo mass
distributions of these subsamples. 

\begin{figure*}
\includegraphics[width=0.5\textwidth]{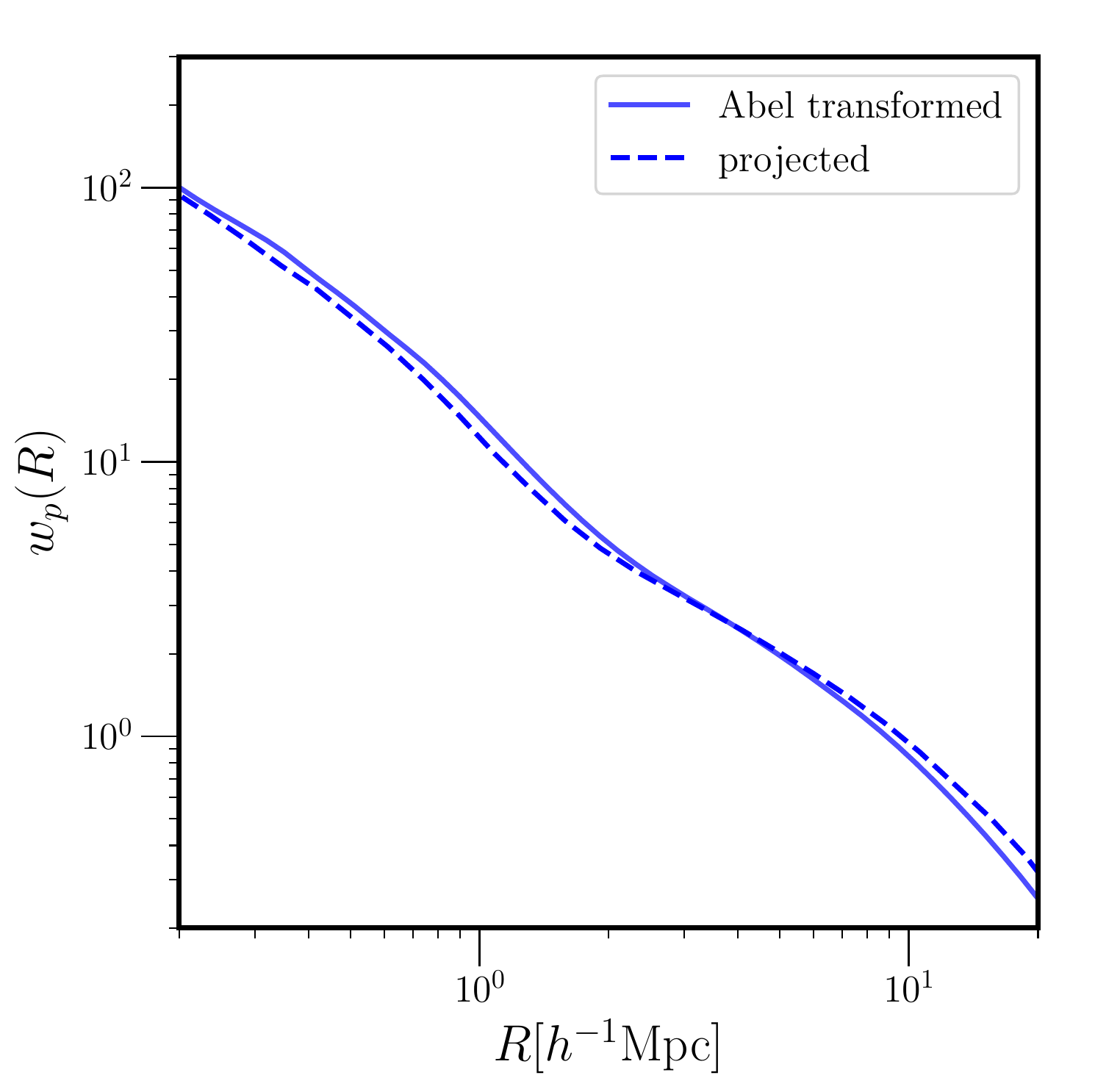}\includegraphics[width=0.5\textwidth]{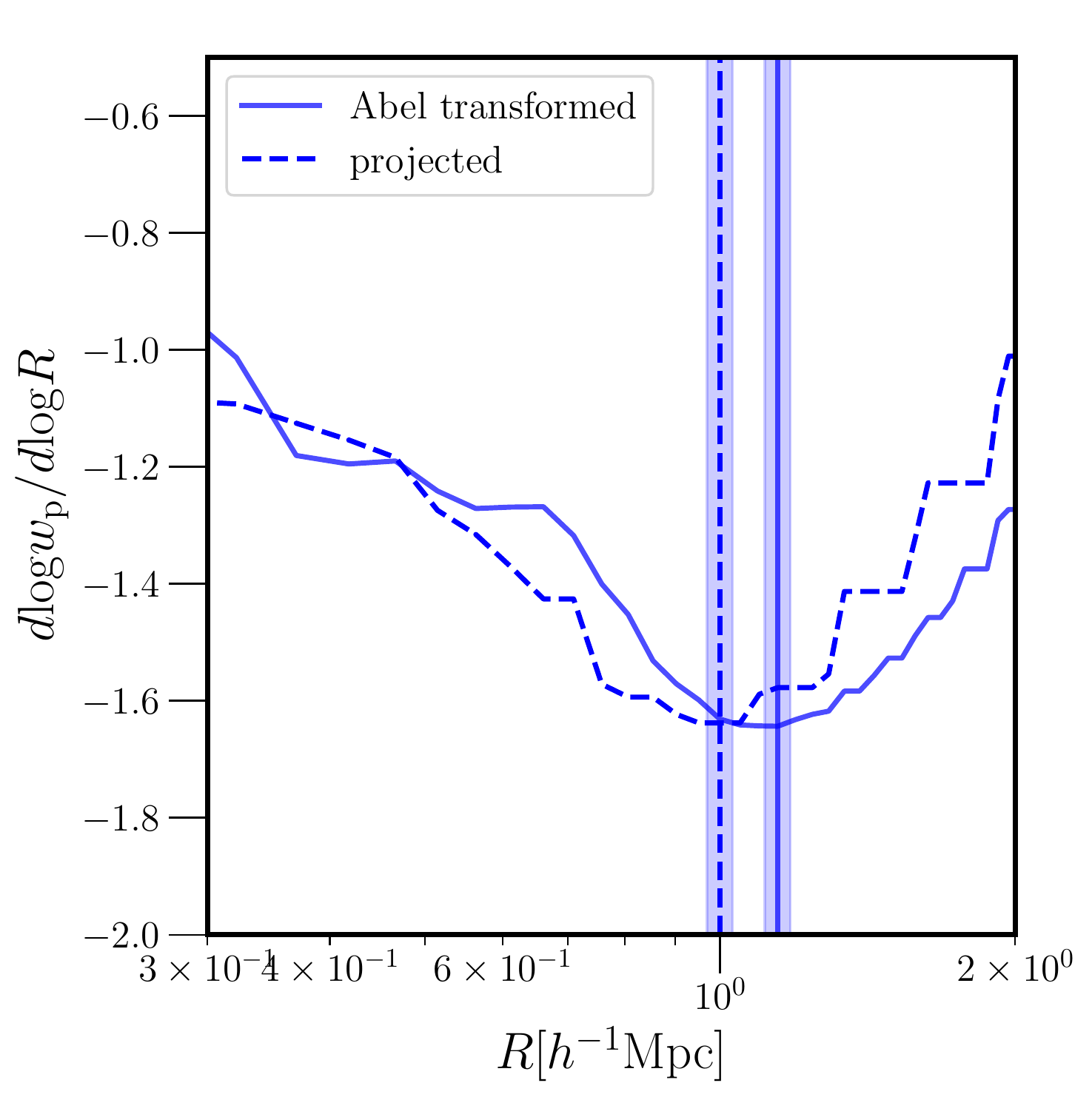}

\caption{\label{fig:Abel}Left: The projected correlation functions transformed
from the three-dimensional cross-correlation functions using Abel transform.
Note that the three-dimensional cross-correlation functions are computed in
real-space. The dashed lines are the projected cross-correlation functions.
Right: Logarithmic derivative profiles of the Abel transformed correlation
functions (solid) as well as the projected correlation functions (dashed). The 
lines correspond to the location of splashback radius, and the shaded regions 
show the typical errors on the measurements in real data taken from \citet{More2016}.
The Abel transformed correlation functions do not agree with the
projected ones because spherical symmetry is broken due to projection effects.}

\end{figure*}

Next, we assume spherical symmetry and project $\xi_{\rm cg}(r)$ to predict the
projected correlation functions using Abel transformation.  If the spherical
symmetry assumption holds then these predictions should match the directly
measured projected correlations. The comparison in the left panel of
Fig.~\ref{fig:Abel}, shows that these two do not agree each other.
This is because spherical symmetry is broken for the clusters
identified by our implementation of redMaPPer algorithm. This is more
evident in the right panel of Fig.\ref{fig:Abel}, which compares the
logarithmic derivative of these correlation functions. The difference in the 
location of the splashback radius between the projected correlation function and the one
from $\xi_{\rm cg}(r)$ is about $15\%$. This difference is suspiciously close
to the size of the difference found between the splashback radius of the
redMaPPer galaxy clusters and the expectation from \lcdm about the location
of their splashback radius \citet{More2016, Baxter2017, Chang2018}. Galaxy
clusters selected using the Sunyaev-Zel'dovich (SZ) effect or via their X-ray
emission are expected to be less severely affected by such effects, however the
comparisons of the measurements of splashback radius to their expected
locations are still dominated by statistical error (e.g.,
\citet{UmetsuDiemer2017, Shin2018, ZurcherMore2019, Contigiani2019}). This also
implies that the splashback radius measured from the projected correlation
functions is not reliable when the clusters in the sample all have a coherent
axis of asymmetry.

\subsection{Comparison with Observations}

We now use data from the SDSS redMaPPer cluster catalog and SDSS
spectroscopic galaxies and look for features similar to what we see in
the simulation analysis from the previous section. To do this, we use
only those redMaPPer clusters which have spectroscopic redshifts and
cross-correlate them with the LOWZ spectroscopic galaxy catalogs.
Unlike \citet{More2016}, our use of galaxies with spectroscopic
redshift allows us to map our the cross-correlation function as a
function of both projected and line-of-sight separations. This allows
us to evaluate the contamination due to foreground/background
galaxies. In particular, we would like to test the dependence of the
projected cross-correlation function on the line-of-sight integration
scale as in Fig.~\ref{fig:assembly_concen} and
Fig.~\ref{fig:assembly_s}.

\begin{figure*}
\includegraphics[width=0.45\textwidth]{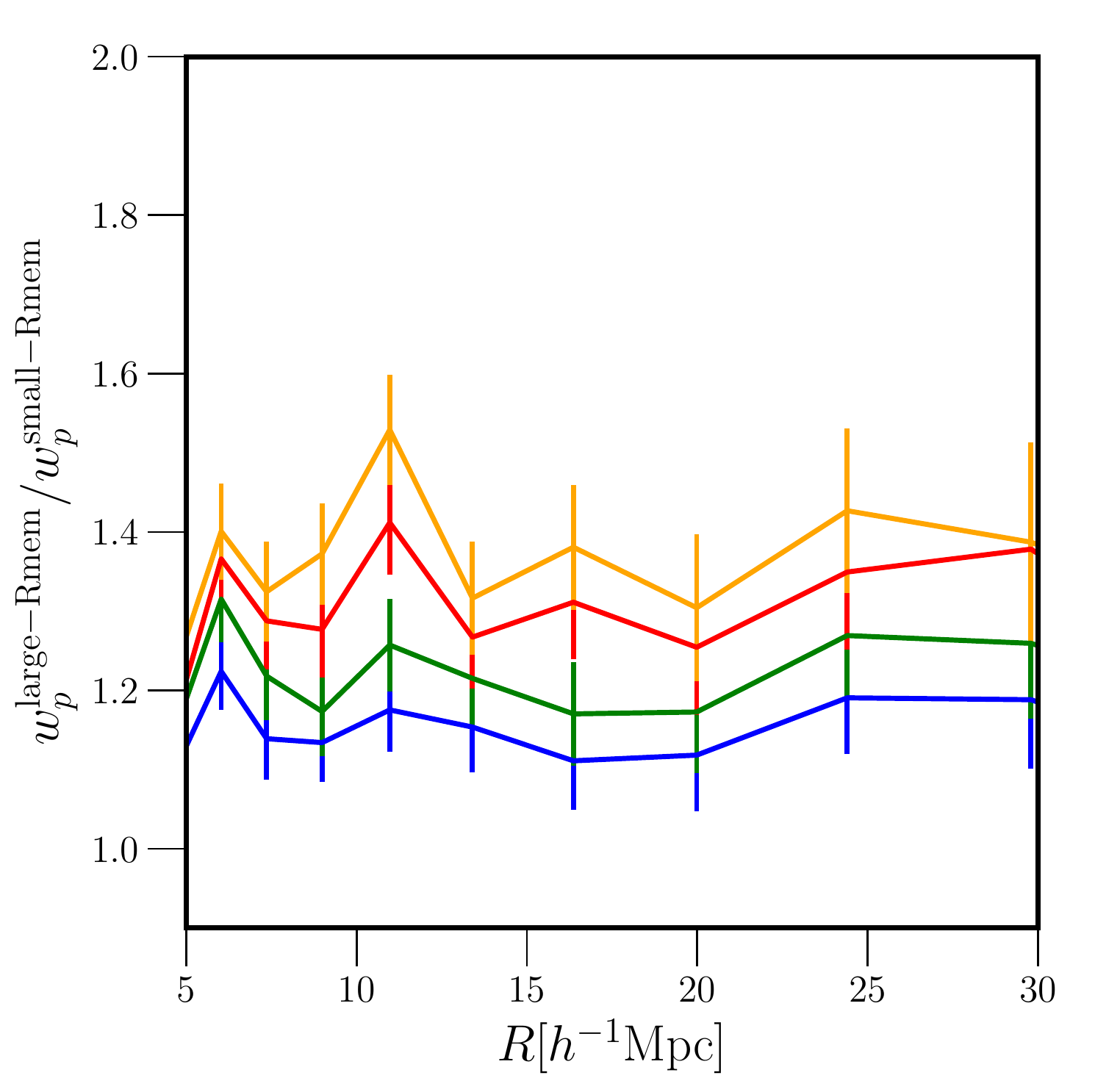}\includegraphics[width=0.45\textwidth]{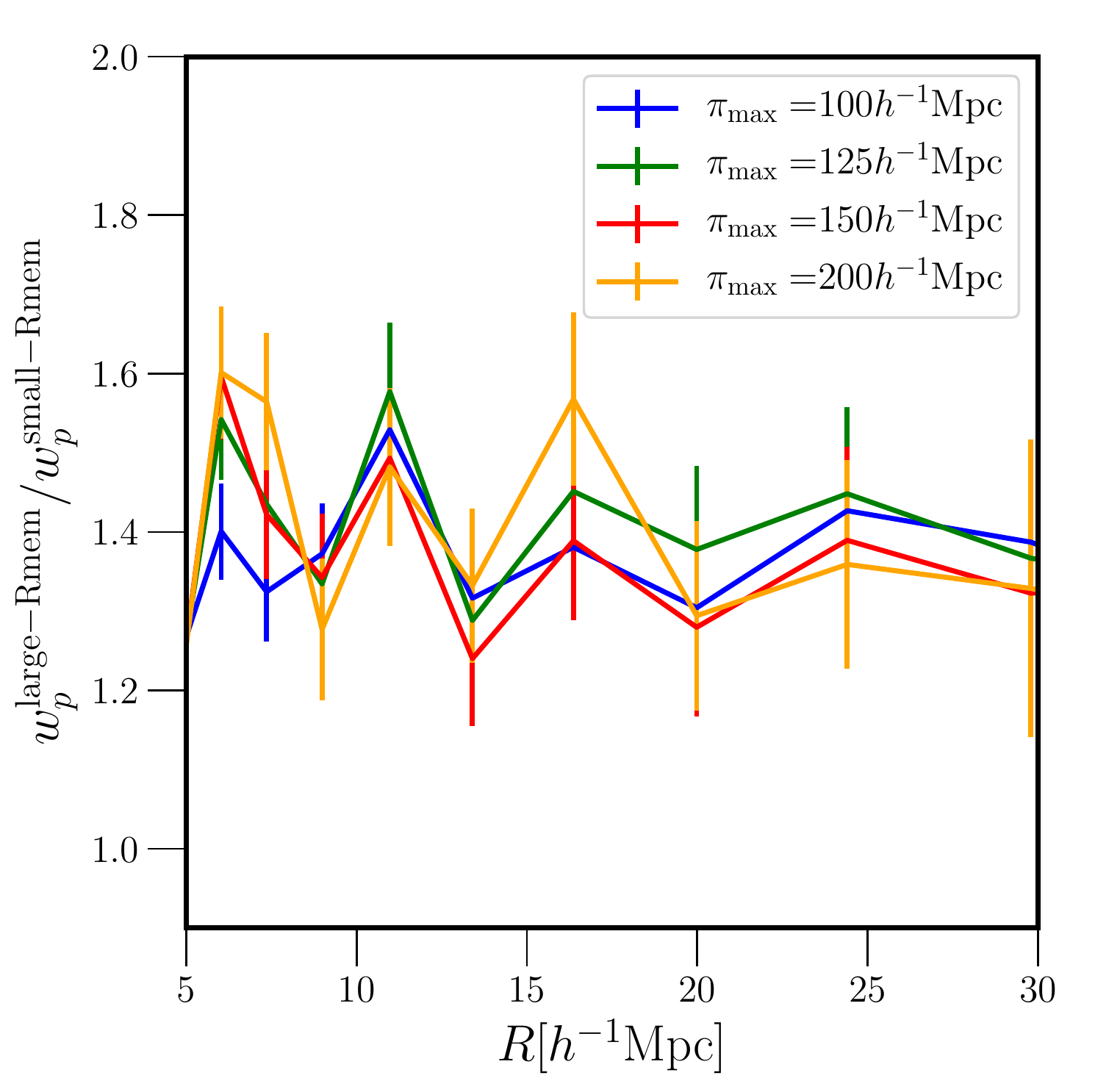}

\caption{\label{fig:assembly-spec}(Left) The ratio of the projected
cross-correlation functions of large-$\avrmem$ and small-$\avrmem$ subsamples
of the SDSS redMaPPer clusters and the LOWZ spectroscopic sample integrated up
to $\pi_{\rm max}=100\mpch$.  Different colors correspond to different integral
scale $R$. As we integrate up to $R=100\mpch$, the ratio becomes larger
compared to the case of $R=10\mpch$ or $R=30\mpch$.  (Right) The same figure
with the integral scale from $\pi_{\rm max}=100\mpch$ to $200\mpch$.  After
exceeding the integral scale $\pi_{\rm max}=100\mpch$, we do not see the
increase in the bias ratio, which implies that the possible projection effect
only goes up to $100\mpch$.}
\end{figure*}

Fig.~\ref{fig:assembly-spec} shows the ratio of the projected
correlation functions between large/small-$\avrmem$ cluster
subsamples. Different colors correspond to different integral scales
$\pi_{\rm max}$ up to $200\mpch$. As is shown in the figure, the ratio
keeps increasing as $\pi_{\rm max}$ increases. In the previous
section, we showed that the ratio does not increase when the integral
scale $\pi_{\rm max}$ becomes larger than the projection length used
to find clusters. Therefore, the figure indicates that the redMaPPer
clusters possibly contain some galaxies which are as far as $100\mpch$
away from their centers. Due to observational limits, it is difficult
to conclude that there are no foreground/background galaxies included
beyond $100\mpch$ even though we do not see any increase in the ratio
beyond $100\mpch$, because the noise becomes dominant beyond a certain
scale.

\section{Summary}

\citet{Miyatake2016} and \citet{More2016} claimed evidence for the
halo assembly bias signal using the redMaPPer galaxy clusters.  These
studies unambiguously show that splitting the sample of the redMaPPer
galaxy clusters into two subsamples based on the compactness of the
member galaxies, leads to samples with similar halo masses, yet
different large scale clustering amplitudes, with large scale biases
of the two subsamples different by almost 60 percent. This difference
was significantly larger than the expected amplitude of the assembly bias
expected from numerical simulations at this mass scale.

\citet{BuschWhite2017} demonstrated that the large assembly bias
signal can be reproduced in a \lcdm framework by mimicking the
optical cluster selection effects. This raised the prospect that the
observational detection of assembly bias could be entirely a result of
projection effects. We conducted a thorough investigation of the
matter using an improved version of their implementation of the mock
redMaPPer algorithm. Following is a succinct summary of our
investigations and findings.

\begin{itemize}
\item We used a novel method in which we applied a redMaPPer like
algorithm to the Millennium simulation semi-analytical galaxy
catalog as well as a shuffled version of this catalog, where the
signature of halo assembly bias was explicitly erased out.
\item We showed that the our mock version of the redMaPPer algorithm
shows features which are similar to those of the real version of
redMaPPer in terms of the mass-richness relation, as well as the
distribution of halo masses at fixed richness.
\item We found that both the shuffled and the non-shuffled versions of
the catalog, when split in to two subsamples each based on
cluster-centric distances at fixed richness, show differences in
clustering amplitude consistent with each other.
\item Given that the shuffled catalog, has no inherent assembly bias
signal, the difference in the observed clustering amplitude of the
subsamples, thus, cannot be used to claim a detection of halo assembly
bias signal.
\item We attributed the presence of the clustering difference to the
correlation induced by the optical cluster selection between the
$\avrmem$ of galaxy clusters, the interloper fractions, and the large
scale structure overdensity modes along the line-of-sight.
\item We also critically investigated the inference of the 3-d
splashback radius of optically selected galaxy clusters based on the
2-d projected density profiles of galaxies around such clusters.
\item We explicitly show that the asymmetry introduced by the
line-of-sight projection effects in optically selected galaxy clusters
hinders a straightforward inference of the 3-d splashback radius of
galaxy clusters.
\item We further verified that the SDSS redMaPPer clusters show the
same dependence of the assembly bias signal on the line-of-sight
projection length as seen in the simulations, further building
circumstantial evidence for the existence of projection effects in
redMaPPer optical clusters.
\end{itemize}

\section{Acknowledgements}
We thank useful discussions with Phillip Busch, Neal Dalal, Benedikt Diemer,
 Bhuvnesh Jain, Andrey Kravtsov, Yen-ting Lin, Hironao Miyatake, Ryoma Murata, 
 Masamune Oguri, Masahiro Takada, Simon White, Ying Zu, and the organizers and 
 participants of the KITP program: The Galaxy-Halo Connection. KITP is supported 
 by the National Science Foundation underGrant No. NSF PHY17-48958. TS is supported
 by Kakenhi grant 15H05893 and 15K21733. SM is supported by Kakenhi grant 16H01089.

\section*{AppendixA: 2D correlation functions}
\begin{figure*}
\includegraphics[width=0.45\textwidth]{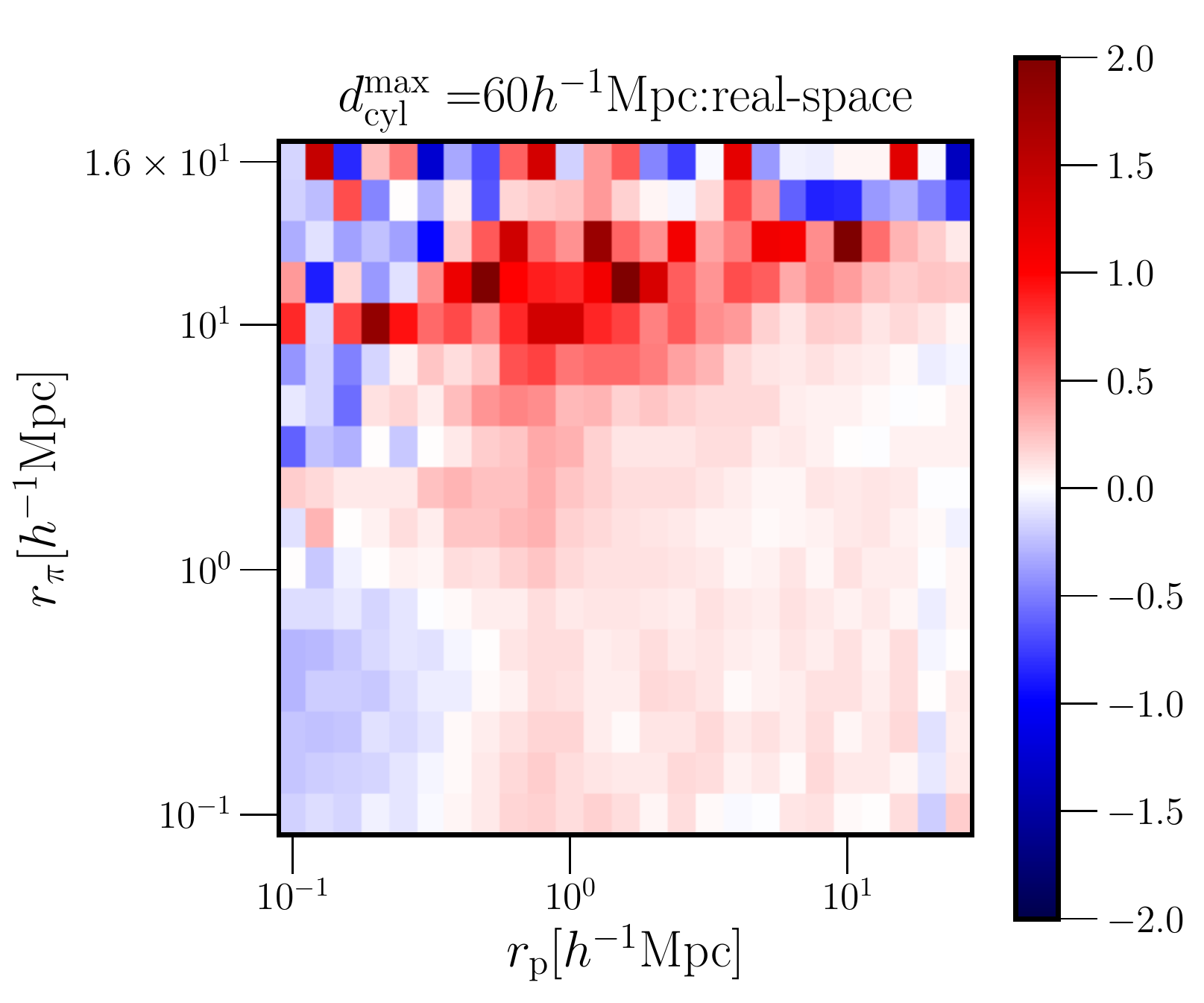}\includegraphics[width=0.45\textwidth]{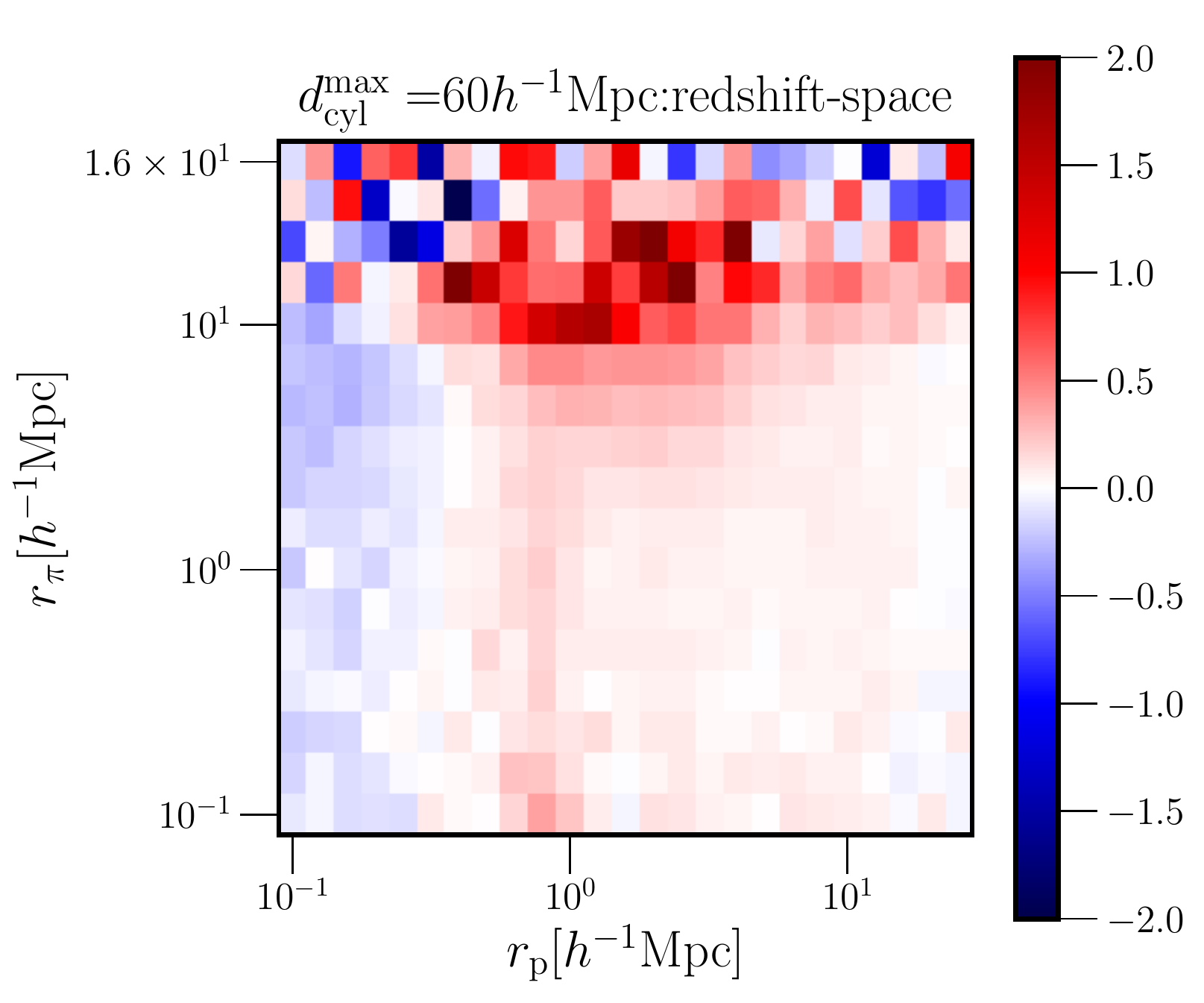}
\includegraphics[width=0.45\textwidth]{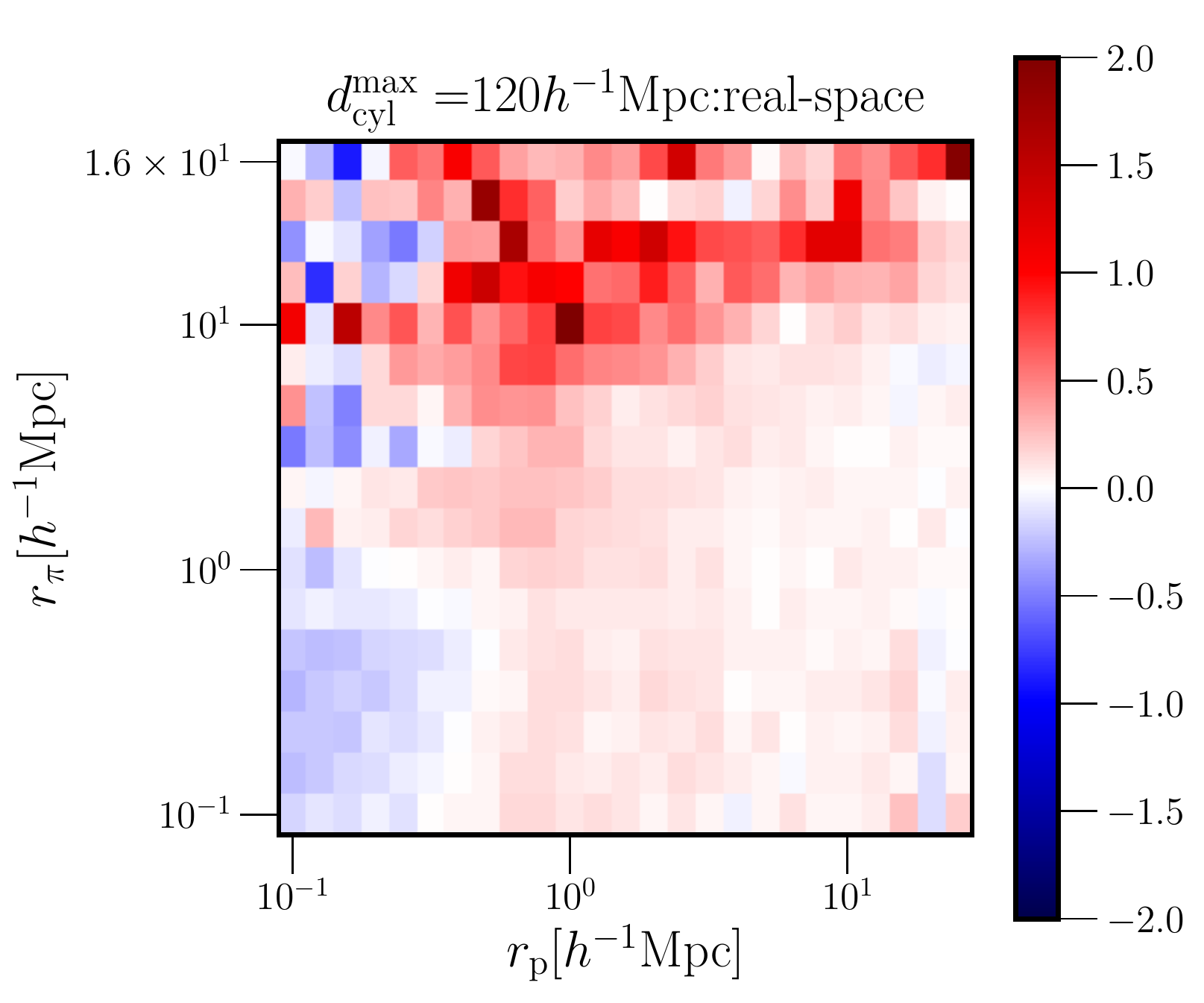}\includegraphics[width=0.45\textwidth]{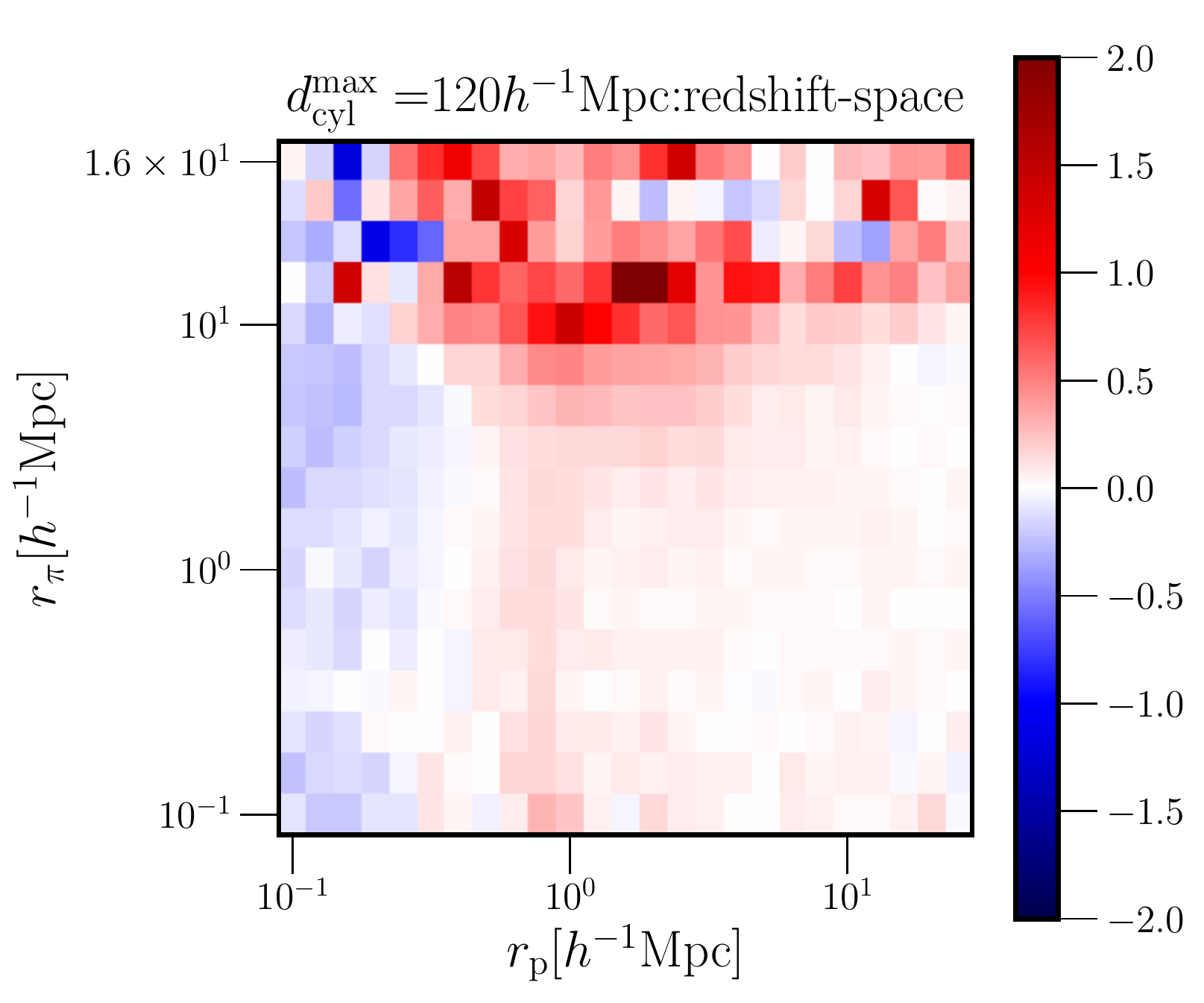}
\includegraphics[width=0.45\textwidth]{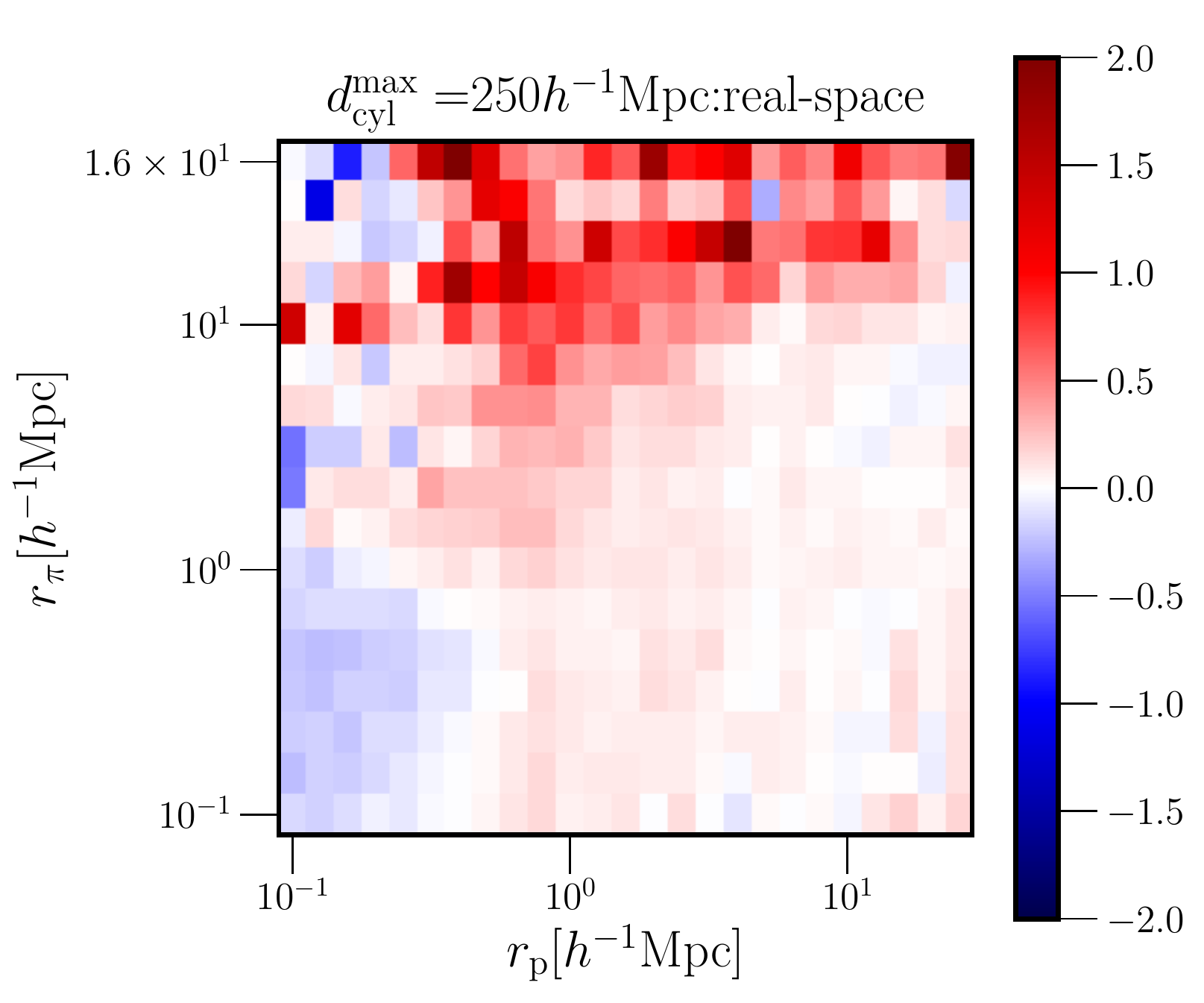}\includegraphics[width=0.45\textwidth]{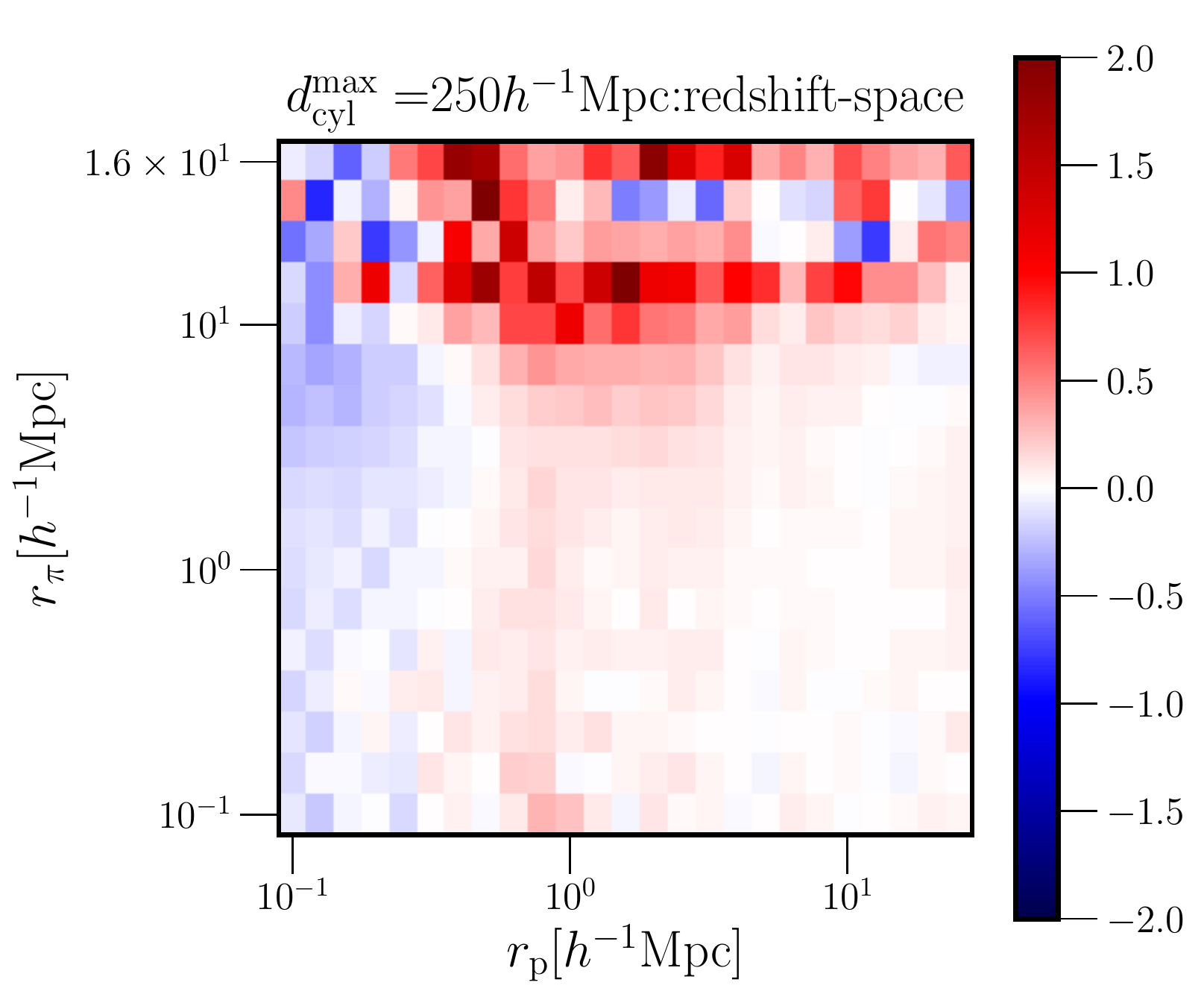}
\caption{\label{fig:2D_corr} The logarithmic ratio of 2D cross-correlation functions between galaxies and the 
large/small-$<R_{\rm mem}>$ cluster subsamples, ${\rm log}_{10}(b_{\rm large-R}/b_{\rm small-R})$. The top to bottom panels correspond to different projection lengths 
$d_{\rm cyl}^{\rm max}=60\mpch$, $120\mpch$, and $250\mpch$ respectively. The left column is 
the ratios computed in real-space, while the right column is the ratios in redshift-space. As is consistent with \citet{BuschWhite2017},
the boundary is smeared out by redshift-space distortions.}

\end{figure*}

In this paper, we discussed how projection effects differently affect
the apparent assembly bias and splashback radius signals between
large/small-$<R_{\rm mem}>$ subsamples. In this Appendix, we compare
the two-dimensional cross correlation functions between the two
subsamples. Fig.~\ref{fig:2D_corr} shows the logarithmic ratio of 2D
cross-correlation functions, ${\rm log}_{10}(b_{\rm large-R}/b_{\rm
small-R})$ for various projection lengths.  In the inner region, the
clustering of small-$R_{\rm mem}$ subsample is stronger, and only goes
to $1\mpch$ in real-space while it extends to $10\mpch$ in
redshift-space. This trend is consistent with \citet{BuschWhite2017}.
On large scales, the behaviour switches and the clustering of
large-$\avrmem$ subsample becomes stronger.

\bibliographystyle{mn2e}
\bibliography{VmaxMvir,redmapper}

\begin{thebibliography}{47}
\expandafter\ifx\csname natexlab\endcsname\relax\def\natexlab#1{#1}\fi

\bibitem[{Adhikari {et~al}\mbox{.}(2014)Adhikari, Dalal, \&
  Chamberlain}]{adhikari2014splashback}
Adhikari S., Dalal N., Chamberlain R.~T., 2014, Journal of Cosmology and
  Astroparticle Physics, 2014, 019

\bibitem[{{Aihara}(2011)}]{Aihara_etal2011}
{Aihara} H. e.~a., 2011, \apjs, 193, 29

\bibitem[{{Alam} {et~al}\mbox{.}(2015){Alam}, {Albareti}, {Allende Prieto},
  {Anders}, {Anderson}, {Anderton}, {Andrews}, {Armengaud}, {Aubourg},
  {Bailey}, \& et~al.}]{Alam2015}
{Alam} S. {et~al.}, 2015, \apjs, 219, 12

\bibitem[{{Bartelmann}(1996)}]{Bartelmann96}
{Bartelmann} M., 1996, \aap, 313, 697

\bibitem[{{Baxter} {et~al}\mbox{.}(2017){Baxter}, {Chang}, {Jain}, {Adhikari},
  {Dalal}, {Kravtsov}, {More}, {Rozo}, {Rykoff}, \& {Sheth}}]{Baxter2017}
{Baxter} E. {et~al.}, 2017, \apj, 841, 18

\bibitem[{{Busch} \& {White}(2017)}]{BuschWhite2017}
{Busch} P., {White} S.~D.~M., 2017, \mnras, 470, 4767

\bibitem[{{Chang} {et~al}\mbox{.}(2018){Chang}, {Baxter}, {Jain},
  {S{\'a}nchez}, {Adhikari}, {Varga}, {Fang}, {Rozo}, {Rykoff}, {Kravtsov},
  {Gruen}, {Hartley}, {Huff}, {Jarvis}, {Kim}, {Prat}, {MacCrann},
  {McClintock}, {Palmese}, {Rapetti}, {Rollins}, {Samuroff}, {Sheldon},
  {Troxel}, {Wechsler}, {Zhang}, {Zuntz}, {Abbott}, {Abdalla}, {Allam},
  {Annis}, {Bechtol}, {Benoit-L{\'e}vy}, {Bernstein}, {Brooks}, {Buckley-Geer},
  {Carnero Rosell}, {Carrasco Kind}, {Carretero}, {D'Andrea}, {da Costa},
  {Davis}, {Desai}, {Diehl}, {Dietrich}, {Drlica-Wagner}, {Eifler}, {Flaugher},
  {Fosalba}, {Frieman}, {Garc{\'\i}a-Bellido}, {Gaztanaga}, {Gerdes},
  {Gruendl}, {Gschwend}, {Gutierrez}, {Honscheid}, {James}, {Jeltema},
  {Krause}, {Kuehn}, {Lahav}, {Lima}, {March}, {Marshall}, {Martini},
  {Melchior}, {Menanteau}, {Miquel}, {Mohr}, {Nord}, {Ogando}, {Plazas},
  {Sanchez}, {Scarpine}, {Schindler}, {Schubnell}, {Sevilla-Noarbe}, {Smith},
  {Smith}, {Soares-Santos}, {Sobreira}, {Suchyta}, {Swanson}, {Tarle},
  {Weller}, \& {DES Collaboration}}]{Chang2018}
{Chang} C. {et~al.}, 2018, \apj, 864, 83

\bibitem[{{Contigiani} {et~al}\mbox{.}(2019){Contigiani}, {Hoekstra}, \&
  {Bah{\'e}}}]{Contigiani2019}
{Contigiani} O., {Hoekstra} H., {Bah{\'e}} Y.~M., 2019, \mnras, 485, 408

\bibitem[{{Dalal} {et~al}\mbox{.}(2008){Dalal}, {White}, {Bond}, \&
  {Shirokov}}]{dalal_etal08}
{Dalal} N., {White} M., {Bond} J.~R., {Shirokov} A., 2008, \apj, 687, 12

\bibitem[{{Diemer} \& {Kravtsov}(2014{\natexlab{a}})}]{diemer14}
{Diemer} B., {Kravtsov} A.~V., 2014{\natexlab{a}}, \apj, 789, 1

\bibitem[{{Diemer} \& {Kravtsov}(2014{\natexlab{b}})}]{DiemerKravtsov2014}
{Diemer} B., {Kravtsov} A.~V., 2014{\natexlab{b}}, \apj, 789, 1

\bibitem[{Diemer {et~al}\mbox{.}(2017)Diemer, Mansfield, Kravtsov, \&
  More}]{diemer2017splashback}
Diemer B., Mansfield P., Kravtsov A.~V., More S., 2017, The Astrophysical
  Journal, 843, 140

\bibitem[{{Efstathiou} {et~al}\mbox{.}(1988){Efstathiou}, {Frenk}, {White}, \&
  {Davis}}]{Efstathiou1988}
{Efstathiou} G., {Frenk} C.~S., {White} S.~D.~M., {Davis} M., 1988, \mnras,
  235, 715

\bibitem[{{Faltenbacher} \& {White}(2010)}]{faltenbacher_white10}
{Faltenbacher} A., {White} S.~D.~M., 2010, \apj, 708, 469

\bibitem[{{Gao} {et~al}\mbox{.}(2005){Gao}, {Springel}, \&
  {White}}]{gao_etal05}
{Gao} L., {Springel} V., {White} S.~D.~M., 2005, \mnras, 363, L66

\bibitem[{{Gao} \& {White}(2007)}]{gao_white07}
{Gao} L., {White} S.~D.~M., 2007, \mnras, 377, L5

\bibitem[{{Guo} {et~al}\mbox{.}(2011){Guo}, {White}, {Boylan-Kolchin}, {De
  Lucia}, {Kauffmann}, {Lemson}, {Li}, {Springel}, \& {Weinmann}}]{guo_etal11b}
{Guo} Q. {et~al.}, 2011, \mnras, 413, 101

\bibitem[{{Han} {et~al}\mbox{.}(2018){Han}, {Cole}, {Frenk}, {Benitez-Llambay},
  \& {Helly}}]{Han_etal2018}
{Han} J., {Cole} S., {Frenk} C.~S., {Benitez-Llambay} A., {Helly} J., 2018,
  \mnras, 474, 604

\bibitem[{{Kaiser}(1984)}]{Kaiser1984}
{Kaiser} N., 1984, \apjl, 284, L9

\bibitem[{{Klypin} {et~al}\mbox{.}(2016){Klypin}, {Yepes}, {Gottl{\"o}ber},
  {Prada}, \& {He{\ss}}}]{Klypin_etal2016}
{Klypin} A., {Yepes} G., {Gottl{\"o}ber} S., {Prada} F., {He{\ss}} S., 2016,
  \mnras, 457, 4340

\bibitem[{{Lin} {et~al}\mbox{.}(2016){Lin}, {Mandelbaum}, {Huang}, {Huang},
  {Dalal}, {Diemer}, {Jian}, \& {Kravtsov}}]{Lin2016}
{Lin} Y.-T., {Mandelbaum} R., {Huang} Y.-H., {Huang} H.-J., {Dalal} N.,
  {Diemer} B., {Jian} H.-Y., {Kravtsov} A., 2016, \apj, 819, 119

\bibitem[{Mansfield {et~al}\mbox{.}(2017)Mansfield, Kravtsov, \&
  Diemer}]{mansfield2017splashback}
Mansfield P., Kravtsov A.~V., Diemer B., 2017, The Astrophysical Journal, 841,
  34

\bibitem[{{Miyatake} {et~al}\mbox{.}(2016){Miyatake}, {More}, {Takada},
  {Spergel}, {Mandelbaum}, {Rykoff}, \& {Rozo}}]{Miyatake2016}
{Miyatake} H., {More} S., {Takada} M., {Spergel} D.~N., {Mandelbaum} R.,
  {Rykoff} E.~S., {Rozo} E., 2016, Physical Review Letters, 116, 041301

\bibitem[{{Mo} \& {White}(1996)}]{MoWhite1996}
{Mo} H.~J., {White} S.~D.~M., 1996, \mnras, 282, 347

\bibitem[{{More}(2016)}]{MoreASLcode}
{More} S., 2016, {SavGolFilterCov: Savitzky Golay filter for data with error
  covariance}. Astrophysics Source Code Library

\bibitem[{{More} {et~al}\mbox{.}(2015{\natexlab{a}}){More}, {Diemer}, \&
  {Kravtsov}}]{more15}
{More} S., {Diemer} B., {Kravtsov} A., 2015{\natexlab{a}}, ArXiv e-prints

\bibitem[{{More} {et~al}\mbox{.}(2015{\natexlab{b}}){More}, {Diemer}, \&
  {Kravtsov}}]{MDK2015}
{More} S., {Diemer} B., {Kravtsov} A.~V., 2015{\natexlab{b}}, \apj, 810, 36

\bibitem[{{More} {et~al}\mbox{.}(2016){More}, {Miyatake}, {Takada}, {Diemer},
  {Kravtsov}, {Dalal}, {More}, {Murata}, {Mandelbaum}, {Rozo}, {Rykoff},
  {Oguri}, \& {Spergel}}]{More2016}
{More} S. {et~al.}, 2016, \apj, 825, 39

\bibitem[{{Murata} {et~al}\mbox{.}(2018){Murata}, {Nishimichi}, {Takada},
  {Miyatake}, {Shirasaki}, {More}, {Takahashi}, \& {Osato}}]{Murata_etal2018}
{Murata} R., {Nishimichi} T., {Takada} M., {Miyatake} H., {Shirasaki} M.,
  {More} S., {Takahashi} R., {Osato} K., 2018, \apj, 854, 120

\bibitem[{{Navarro} {et~al}\mbox{.}(1997){Navarro}, {Frenk}, \&
  {White}}]{NFW97}
{Navarro} J.~F., {Frenk} C.~S., {White} S.~D.~M., 1997, \apj, 490, 493

\bibitem[{{Rozo} \& {Rykoff}(2014)}]{Rozo2014}
{Rozo} E., {Rykoff} E.~S., 2014, \apj, 783, 80

\bibitem[{{Rozo} {et~al}\mbox{.}(2015{\natexlab{a}}){Rozo}, {Rykoff},
  {Bartlett}, \& {Melin}}]{Rozo2015}
{Rozo} E., {Rykoff} E.~S., {Bartlett} J.~G., {Melin} J.-B., 2015{\natexlab{a}},
  \mnras, 450, 592

\bibitem[{{Rozo} {et~al}\mbox{.}(2015{\natexlab{b}}){Rozo}, {Rykoff}, {Becker},
  {Reddick}, \& {Wechsler}}]{Rozo2015_2}
{Rozo} E., {Rykoff} E.~S., {Becker} M., {Reddick} R.~M., {Wechsler} R.~H.,
  2015{\natexlab{b}}, \mnras, 453, 38

\bibitem[{{Rykoff} {et~al}\mbox{.}(2012){Rykoff}, {Koester}, {Rozo}, {Annis},
  {Evrard}, {Hansen}, {Hao}, {Johnston}, {McKay}, \& {Wechsler}}]{Rykoff2012}
{Rykoff} E.~S. {et~al.}, 2012, \apj, 746, 178

\bibitem[{{Rykoff} {et~al}\mbox{.}(2014){Rykoff}, {Rozo}, {Busha}, {Cunha},
  {Finoguenov}, {Evrard}, {Hao}, {Koester}, {Leauthaud}, {Nord}, {Pierre},
  {Reddick}, {Sadibekova}, {Sheldon}, \& {Wechsler}}]{Rykoff_etal2014}
{Rykoff} E.~S. {et~al.}, 2014, \apj, 785, 104

\bibitem[{Shi(2016)}]{shi2016outer}
Shi X., 2016, Monthly Notices of the Royal Astronomical Society, 459, 3711

\bibitem[{{Shin} {et~al}\mbox{.}(2018){Shin}, {Adhikari}, {Baxter}, {Chang},
  {Jain}, {Battaglia}, {Bleem}, {Bocquet}, {DeRose}, {Gruen}, {Hilton},
  {Kravtsov}, {McClintock}, {Rozo}, {Rykoff}, {Varga}, {Wechsler}, {Wu},
  {Aiola}, {Allam}, {Bechtol}, {Benson}, {Bertin}, {Bond}, {Brodwin}, {Brooks},
  {Buckley-Geer}, {Burke}, {Carlstrom}, {Carnero Rosell}, {Carrasco Kind},
  {Carretero}, {Castander}, {Choi}, {Cunha}, {Crawford}, {da Costa}, {De
  Vicente}, {Desai}, {Devlin}, {Dietrich}, {Doel}, {Dunkley}, {Eifler},
  {Evrard}, {Flaugher}, {Fosalba}, {Gallardo}, {Garc{\'{\i}}a-Bellido},
  {Gaztanaga}, {Gerdes}, {Gralla}, {Gruendl}, {Gschwend}, {Gupta}, {Gutierrez},
  {Hartley}, {Hill}, {Ho}, {Hollowood}, {Honscheid}, {Hoyle}, {Huffenberger},
  {Hughes}, {James}, {Jeltema}, {Kim}, {Krause}, {Kuehn}, {Lahav}, {Lima},
  {Madhavacheril}, {Maia}, {Marshall}, {Maurin}, {McMahon}, {Menanteau},
  {Miller}, {Miquel}, {Mohr}, {Naess}, {Nati}, {Newburgh}, {Niemack}, {Ogando},
  {Partridge}, {Patil}, {Plazas}, {Rapetti}, {Reichardt}, {Romer}, {Sanchez},
  {Scarpine}, {Schindler}, {Serrano}, {Smith}, {Smith}, {Soares-Santos},
  {Sobreira}, {Staggs}, {Stark}, {Stein}, {Suchyta}, {Swanson}, {Tarle},
  {Thomas}, {van Engelen}, {Wollack}, {Xu}, \& {Zhang}}]{Shin2018}
{Shin} T. {et~al.}, 2018, arXiv e-prints

\bibitem[{{Springel} {et~al}\mbox{.}(2005){Springel}, {White}, {Jenkins},
  {Frenk}, {Yoshida}, {Gao}, {Navarro}, {Thacker}, {Croton}, {Helly},
  {Peacock}, {Cole}, {Thomas}, {Couchman}, {Evrard}, {Colberg}, \&
  {Pearce}}]{springel_etal05}
{Springel} V. {et~al.}, 2005, \nat, 435, 629

\bibitem[{{Springel} {et~al}\mbox{.}(2001){Springel}, {White}, {Tormen}, \&
  {Kauffmann}}]{springel2001}
{Springel} V., {White} S.~D.~M., {Tormen} G., {Kauffmann} G., 2001, \mnras,
  328, 726

\bibitem[{{Umetsu} \& {Diemer}(2017)}]{UmetsuDiemer2017}
{Umetsu} K., {Diemer} B., 2017, \apj, 836, 231

\bibitem[{{van den Bosch} \& {Ogiya}(2018)}]{vdBosch_etal2018b}
{van den Bosch} F.~C., {Ogiya} G., 2018, \mnras, 475, 4066

\bibitem[{{van den Bosch} {et~al}\mbox{.}(2018){van den Bosch}, {Ogiya},
  {Hahn}, \& {Burkert}}]{vdBosch_etal2018a}
{van den Bosch} F.~C., {Ogiya} G., {Hahn} O., {Burkert} A., 2018, \mnras, 474,
  3043

\bibitem[{{Wang} {et~al}\mbox{.}(2013){Wang}, {Weinmann}, {De Lucia}, \&
  {Yang}}]{wang_etal13}
{Wang} L., {Weinmann} S.~M., {De Lucia} G., {Yang} X., 2013, \mnras, 433, 515

\bibitem[{{Wechsler} {et~al}\mbox{.}(2006){Wechsler}, {Zentner}, {Bullock},
  {Kravtsov}, \& {Allgood}}]{wechsler06}
{Wechsler} R.~H., {Zentner} A.~R., {Bullock} J.~S., {Kravtsov} A.~V., {Allgood}
  B., 2006, \apj, 652, 71

\bibitem[{{Yang} {et~al}\mbox{.}(2006){Yang}, {Mo}, \& {van den
  Bosch}}]{Yang2006}
{Yang} X., {Mo} H.~J., {van den Bosch} F.~C., 2006, \apjl, 638, L55

\bibitem[{{Zu} {et~al}\mbox{.}(2017){Zu}, {Mandelbaum}, {Simet}, {Rozo}, \&
  {Rykoff}}]{Zu2017}
{Zu} Y., {Mandelbaum} R., {Simet} M., {Rozo} E., {Rykoff} E.~S., 2017, \mnras,
  470, 551

\bibitem[{{Z{\"u}rcher} \& {More}(2019)}]{ZurcherMore2019}
{Z{\"u}rcher} D., {More} S., 2019, \apj, 874, 184

\end{thebibliography}

\end{document}